\documentclass[prl,aps,superscriptaddress,footinbib,twocolumn]{revtex4-2}\usepackage[latin9]{inputenc}
\usepackage{color}
\usepackage{amsmath}
\usepackage{amsthm}
\usepackage{amssymb}
\usepackage{stmaryrd}
\usepackage{graphicx}
\usepackage[unicode=true,
 bookmarks=true,bookmarksnumbered=false,bookmarksopen=false,
 breaklinks=false,pdfborder={0 0 1},backref=false,colorlinks=true]
 {hyperref}
\hypersetup{
 linkcolor=magenta, urlcolor=blue, citecolor=blue, pdfstartview={FitH}, hyperfootnotes=false, unicode=true}
\makeatletter

\usepackage{amsfonts}\usepackage{tabularx}\usepackage{dcolumn}\usepackage{bm}\usepackage{graphicx}\usepackage{epstopdf}
\usepackage{times}
\usepackage{cleveref}
\hypersetup{urlcolor=blue}

\def\ket#1{\left|#1\right\rangle}
\def\bra#1{\left\langle#1\right|}
\def\braket#1{\left\langle#1\right\rangle}
\def\ketbra#1{\left|#1\right\rangle\!\left\langle#1\right|}

\newcommand{\cD}{\mathcal{D}}

\newcommand{\btheta}{\bm{\theta}}
\newcommand{\bx}{\bm{x}}
\newcommand{\E}{\mathbb{E}}
\newcommand{\norm}[1]{\left\lVert#1\right\rVert}
\DeclareMathOperator{\Tr}{Tr}
\newtheorem{theorem}{Theorem}

\makeatother

\begin{document}

\title{Quantum automated learning with provable and explainable trainability}

\author{Qi Ye}\thanks{These authors contributed equally}\affiliation{Center for Quantum Information, IIIS, Tsinghua University, Beijing 100084, China}\affiliation{Shanghai Qi Zhi Institute, Shanghai 200232, China}\affiliation{School of Engineering and Applied Sciences, Harvard University, Cambridge, Massachusetts 02138, USA}
\author{Shuangyue Geng}\thanks{These authors contributed equally}\affiliation{Center for Quantum Information, IIIS, Tsinghua University, Beijing 100084, China}\affiliation{Shanghai Qi Zhi Institute, Shanghai 200232, China}

\author{Zizhao Han}\affiliation{Center for Quantum Information, IIIS, Tsinghua University, Beijing 100084, China}
\author{Weikang Li}
\affiliation{Center for Quantum Information, IIIS, Tsinghua University, Beijing 100084, China}

\author{L.-M. Duan}
\affiliation{Center for Quantum Information, IIIS, Tsinghua University, Beijing 100084, China}
\affiliation{Hefei National Laboratory, Hefei 230088, China}

\author{Dong-Ling Deng}
\email{dldeng@tsinghua.edu.cn}
\affiliation{Center for Quantum Information, IIIS, Tsinghua University, Beijing 100084, China}\affiliation{Shanghai Qi Zhi Institute, Shanghai 200232, China}
\affiliation{Hefei National Laboratory, Hefei 230088, China}

\begin{abstract}
{Machine learning is widely believed to be one of the most promising practical applications of quantum computing. Existing quantum machine learning schemes typically employ a quantum-classical hybrid approach that relies crucially on gradients of model parameters. Such an approach lacks provable convergence to global minima and will become infeasible as quantum learning models scale up. Here, we introduce quantum automated learning, where \textit{no} variational parameter is involved and the training process is converted to quantum state preparation. In particular, we encode training data into unitary operations and iteratively evolve a random initial state under these unitaries and their inverses, with a target-oriented perturbation towards higher prediction accuracy sandwiched in between. Under reasonable assumptions, we rigorously prove that the evolution converges exponentially to the desired state corresponding to the global minimum of the loss function. We show that such a training process can be understood from the perspective of preparing quantum states by imaginary time evolution, where the data-encoded unitaries together with target-oriented perturbations would train the quantum learning model in an automated fashion. We further prove that the introduced quantum automated learning paradigm features good generalization ability with the generalization error upper bounded by the ratio between a logarithmic function of the Hilbert space dimension and the number of training samples. In addition, we carry out extensive numerical simulations on real-life images and quantum data to demonstrate the effectiveness of our approach and validate the assumptions. Our results establish an unconventional quantum learning strategy that is gradient-free with provable and explainable trainability, 
which would be crucial for large-scale practical applications of quantum computing in machine learning scenarios. 
}
\end{abstract}

\maketitle

\noindent \textbf{\large{}Introduction}{\large\par}

\noindent Machine learning, the core of artificial intelligence (AI), has achieved dramatic success~\cite{LeCun2015Deep,Goodfellow2016Deep,Wang2023Scientific}. A number of long-standing challenging problems, such as playing the game of Go~\cite{Silver2016Mastering,Silver2017Mastering}, predicting protein structures~\cite{Senior2020Improved}, and automated theorem proving at the olympiad level~\cite{Trinh2024Solving},  have been cracked in recent years, elevating AI to new scientific heights. Yet, as Moore's law is approaching the end and machine learning models become unprecedentedly large devouring a tremendous amount of resources, 
further development of AI would be subject to the limitations of computational power and energy consumption~\cite{2023AI}. 
Quantum computing promises a potential way out of this dilemma. 

Indeed, parallel to machine learning, the field of quantum computing has also made remarkable progress in the past decades~\cite{Preskill2018Quantum}, with experimental demonstrations of quantum supremacy~\cite{Arute2019Quantum,Zhong2020Quantum,Wu2021Strong} and error correction~\cite{Bluvstein2024Logical,Acharya2024Quantum} marked as the latest milestones. 
A variety of quantum algorithms have been proposed to enhance, speed up or innovate machine learning~\cite{Liu2021Rigorous,Huang2021Power}, giving rise to a new vibrant research frontier of quantum machine learning~\cite{Biamonte2017Quantum,Dunjko2018Machine,DasSarma2019Machine}. Proof-of-principle experiments have been reported with current noisy intermediate-scale quantum devices~\cite{Herrmann2022Realizing,Saggio2021Experimental,Ren2022Experimental,Hu2019Quantum,Huang2021Quantum,Gong2023Quantum,Peters2021Machine}, including these on quantum convolutional networks~\cite{Herrmann2022Realizing},   reinforcement learning~\cite{Saggio2021Experimental},   adversarial learning~\cite{Ren2022Experimental}, federated learning~\cite{Liu2025Practical}, and  continual learning~\cite{Zhang2024Quantum}. 
Most of these quantum learning schemes rely on parametrized quantum circuits and utilize gradient-based approaches for training~\cite{Mitarai2018Quantum}. They face three major difficulties when scaling up. First, in classical deep learning the gradients can be calculated in an efficient and parallel way by the backpropagation algorithm~\cite{Wythoff1993Backpropagation}. Whereas, in quantum scenarios the implementation of the backpropagation algorithm is resource demanding~\cite{Beer2020Training,Pan2023Deep} and quantum gradients are usually obtained one-by-one for each variational parameter~\cite{Mitarai2018Quantum,Schuld2019Evaluating}. This renders large-scale quantum learning impractical, especially for models as large as GPT-4~\cite{OpenAI2024GPT4} and GLaM~\cite{Du2022GLaM} with up to trillions of parameters. Second, quantum landscapes can have exponentially many local minima~\cite{Bittel2021Training} and exhibit the notorious barren plateau phenomenon~\cite{Larocca2024Review,McClean2018Barren,Cerezo2021Cost}, where the gradients vanish exponentially with the system size. Consequently, 
the gradient-based training of variational quantum learning models becomes intrinsically hard as they scale up. Third, there lacks a clear understanding of the gradient-based training dynamics to guide further development of large-scale quantum learning models. 

\begin{figure*}
    \centering
    \includegraphics[width=\linewidth]{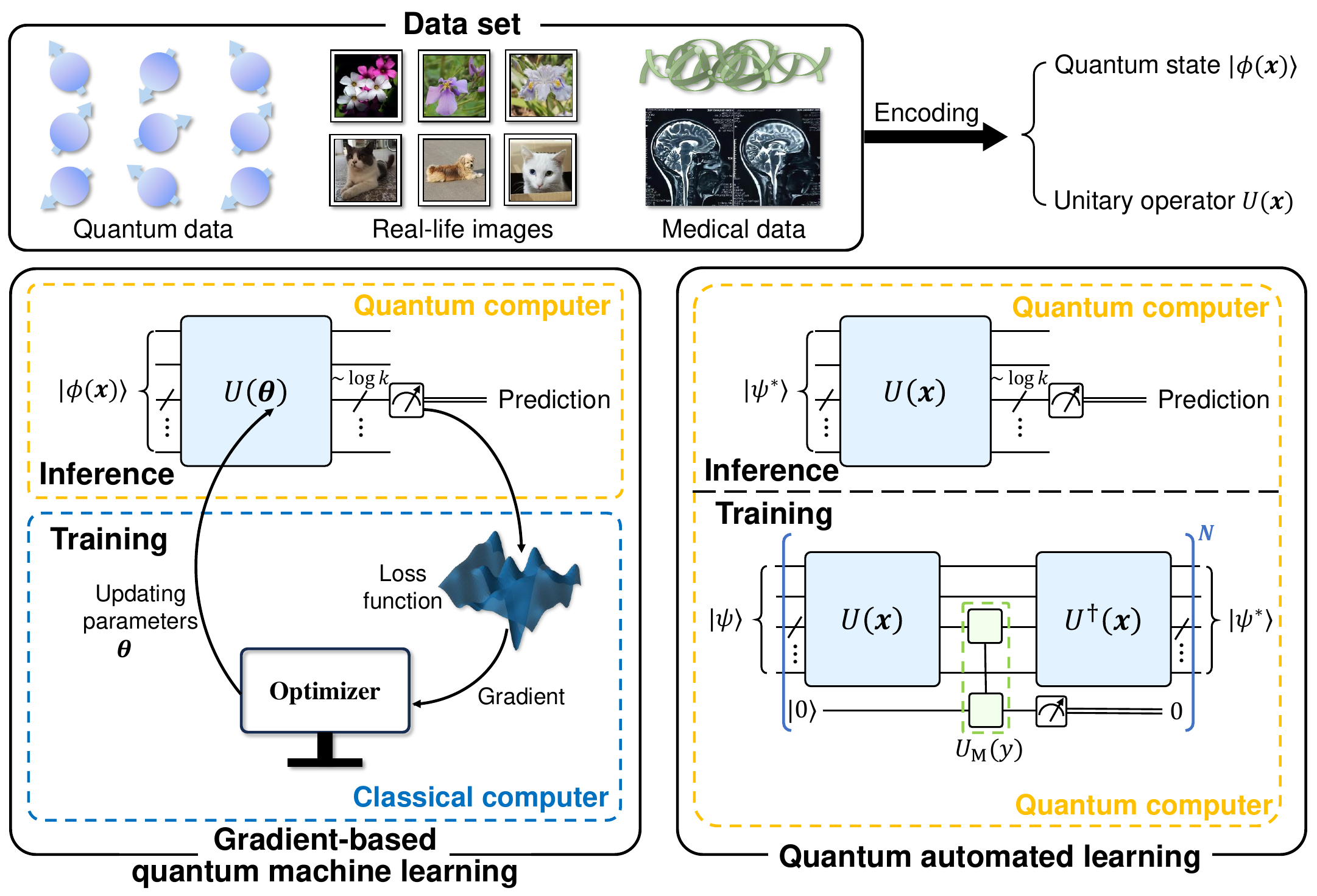}
    \caption{\textbf{Comparison between two different quantum learning paradigms}. Upper panel:  a sketch of quantum  learning and data encoding schemes. Lower left panel: an illustration of gradient-based quantum learning. In this paradigm, the input datum $\bx$ is usually first encoded into a quantum state $\ket{\phi(\bx)}$ and then fed into a parametrized quantum circuit $U(\btheta)$. For a $k$-class classification task, one measures about $\log k$ qubits and obtains the gradient of a predefined loss function on a classical computer based on the measurement results. With the obtained gradient, one updates the variation parameters $\btheta$. This process is iterated a number of times until the loss function no longer decreases (the training process). At the inference stage, one inputs $\ket{\phi(\bx)}$ for an unseen sample $\bx$ into the variational circuit with optimized parameters and then does the same measurement on $\log k$ qubits to make predictions. Lower right panel: a sketch of quantum automated learning. In this scenario,  the input datum $\bx$ is encoded into $U(\bx)$ and the training process is transferred into quantum state preparation, which in turn can be accomplished by a dissipation process with guaranteed convergence. At the inference stage, we evolve the prepared state $\ket{\psi^*}$ with $U(\bx)$ and then do the measurement to make a prediction for the unseen sample $\bx$. The quantum automated learning approach involves no variational parameter and is inherently gradient-free, thus it is more scalable to large-scale practical quantum learning applications.
}
    \label{fig:QAL}
\end{figure*}

In this paper, we circumvent these difficulties by introducing quantum automated learning (QAL) with provable and explainable trainability. Our approach involves \textit{no} variational parameter and thus is inherently gradient-free and scalable (see Fig.~\ref{fig:QAL}). Specifically, we focus our discussion on supervised learning with pre-labeled training data. We convert the training process into quantum state preparation for a target state, on which the outcome of a data-encoded measurement gives the predicted label of the data sample. To prepare the desired state, we design a dissipation process for each training sample that drives a random initial state towards the target state in an automated manner. 
We prove that such a dissipation process converges exponentially to the target state corresponding to the global minimum of the loss function. We show that this training process has a clear physical interpretation: it  essentially implements an imaginary time evolution that cools down
the system to the ground state of a data-encoded Hamiltonian. In addition, we prove that 
the generalization error of our approach is upper bounded by $\sqrt{\log (D) /N}$, where $D$ and $N$ denote the Hilbert space dimension and the number of training samples, respectively. 
To demonstrate
the effectiveness of our approach and validate the assumptions, we carry out extensive numerical simulations on real-life images (e.g., MNIST Dataset~\cite{lecun1998mnist} and Fashion MNIST Dataset~\cite{xiaoFashionMNISTNovelImage2017}) and quantum data (e.g., thermal and localized quantum many-body states). Our QAL strategy eludes the three pronounced challenges faced by conventional gradient-based quantum learning schemes, advancing towards the goal of scaling up quantum learning to large-scale practical applications.

\vspace{.5cm}
\noindent\textbf{\large{}Quantum automated learning}{\large\par}
\noindent We first introduce the general framework for quantum automated learning. For convenience, we focus our discussion on classification tasks in the setting of supervised learning~\cite{Li2022Recent}, where we assign a label $y(\bx)\in Y$ to an input data sample $\bx \in X$, with $Y$ being a finite label set of size $k$ and $X$ the set of all possible samples. We denote the training set as $S_N=\{(\bx_1,y_1),\cdots,(\bx_N,y_N)\}$, where $\bx_i\in X$ is sampled from an unknown distribution $\cD$, $y_i\in Y$ represents the label of $\bx_i$, and $N$ is the size of the training set. The goal of classification is to predict the labels of unseen samples drawn from $\cD$ with high probability.  

A widely studied strategy in solving classification problems relies on quantum neural networks~\cite{Cerezo2021Variational,Li2022Quantum}, where the input datum is first encoded into a quantum state $|\phi (\bx)\rangle$ and then fed into a variational quantum circuit $U(\btheta)$ with parameters collectively denoted by $\btheta$. To train the quantum neural networks, one introduces a loss function that measures the discrepancy between predicted and true labels, and minimizes it on training data via the gradient descent method. The gradients are usually obtained by using finite differences or the parameter-shift rule assisted by a classical computer~\cite{Mitarai2018Quantum,Schuld2019Evaluating}. An illustration of this quantum-classical hybrid approach is sketched in the lower left panel of Fig.~\ref{fig:QAL}. Such an approach is a straightforward extension of classical neural networks to the quantum domain, yet three pronounced difficulties hinder its scalability. The first difficulty concerns the impractical time consumption in obtaining the gradients. Unlike in classical deep learning where the gradients can be calculated in parallel through the backpropagation algorithm, quantum gradients are usually obtained one-by-one for each variational parameter. In fact, even calculating the gradient for a single parameter demands  executing the variational quantum circuit thousands of times. The second difficulty originates from the pathological 
landscapes of quantum neural networks---they typically bear exponentially many local minima and suffer from barren plateaus~\cite{Larocca2024Review,McClean2018Barren,Cerezo2021Cost}. Overcoming this difficulty requires exponentially many runs of the quantum circuit. The third difficulty regards the shortage of a physical understanding of the training dynamics under gradient descent, which precludes a prior guide for tuning hyperparameters or designing more efficient gradient-based algorithms without resorting to trial and error. These three difficulties make the quantum neural network based approach impractical when scaled up, especially for large models with trillions of parameters.

Unlike the gradient-based approach discussed above, quantum automated learning exploits an entirely different strategy that involves no variational parameter and is gradient-free inherently.  As sketched in the lower right panel of Fig.~\ref{fig:QAL},  our general recipe goes as the following: (i) start with a random $n$-qubit initial state $|\psi\rangle$, where $n=O(\log |\bx|)$ 
with $|\bx|$ the dimension of the data sample $\bx$ (Supplementary Sec. II A); (ii) randomly choose a data sample $(\bx,y)$ from the training set and utilize a unitary-encoding scheme to encode $\bx$ into a unitary $U(\bx)$ (Methods and Supplementary Sec. II B); (iii) evolve the system with $U(\bx)$, a $y$-dependent perturbation $M_y$, and $U(\bx)^{\dagger}$; (iv) repeat steps (ii) and (iii) for $T$ times to drive the system to the target state $|\psi^*\rangle$; (v) for an unseen datum $\bx'$, evolve $|\psi^*\rangle$ with $U(\bx')$ and then measure $\sim\log k$ qubits to output the predicted label $y'$. In this scheme, steps (i)-(iv) constitute the training process and step (v) serves as the inference. We stress that such a scheme does not require a quantum random access memory~\cite{Giovannetti2008Quantum}, which is resource-demanding and has not been realized in experiments so far, to load classical data. Indeed, a datum $\bx$ is input into the training process through $U(\bx)$, which consists of multiple layers of single- and two-qubit gates specified by $\bx$. These gates are readily accessible in current experiments. In step (iii), the perturbation $M_y=\ketbra{y}+(1-\eta)(\mathbf{I}-\ketbra{y})$, where $\ketbra{y}$ denotes a $\log k$-qubit state that encodes the label $y$, $\eta$ is the learning rate,  and $\mathbf{I}$ is the identity matrix. This perturbation aims to suppress the probability of incorrect prediction and is implemented by block encoding into a unitary $U_y$ with an ancillary qubit combined with post-selection (Methods and Supplementary Sec. II C). In practice, one may need to compile $U_y$ into hardware-compatible elementary gates. This can be accomplished efficiently by the Solovay-Kitaev algorithm~\cite{Dawson2006TheSolovay} or the quantum compiling algorithm based on reinforcement learning~\cite{Zhang2020Topological} in general. More efficient compiling of $U_y$ exists given its special structure (Supplementary Sec. II C).

From a high-level perspective, the QAL scheme formulates the training process as quantum state preparation, which in turn is accomplished by a data-dependent dissipation process that drives the system to the target state in a fully autonomous way. It naturally circumvents the challenges in designing appropriate quantum neural networks for given tasks and obtaining gradients to train them. When scaling up to large sizes, it is free from local minima and barren plateaus. In addition, owing to the dissipation nature of the ``training'' process, the QAL scheme features some degree of resilience to the noise in the quantum hardware, which makes it particularly suitable for current noisy quantum devices without error correction. One may worry that the success probability of preparing $|\psi^*\rangle$ would become vanishingly small due to post-selection in step (iii) as $T$ increases. This is in fact not the case as proved in Theorem~\ref{thm:analytical} under certain reasonable assumptions and verified in numerical simulations with real-life images. Another plausible (but not true) drawback has a profound relation to the quantum no-cloning theorem~\cite{Wootters1982Single}, which states that it is impossible to copy an arbitrary unknown quantum state. Due to the no-cloning theorem, one cannot make copies of $|\psi^*\rangle$ and hence it seems that for each run of the inference step (v), one has to execute the whole training process (i)-(iv). This is not necessary either, as discussed in-depth later.

\vspace{.5cm}
\noindent \textbf{\large{}Provable convergence and physical explanation}{\large\par}

\noindent We now show that the training process of the QAL scheme introduced above can be understood from the perspective of preparing quantum states through imaginary time evolution and it converges exponentially to the global minimum of a naturally defined loss function. To this end, we define a Hamiltonian $H_{\bx}=\mathbf{I}-U(\bx)^\dagger \Pi_{y(\bx)}U(\bx)$, where $\Pi_{y(\bx)}$  denotes the measurement projection corresponding to the state encoding the label $y(\bx)$. The probability of correct prediction reads $\braket{\psi|U(\bx)^{\dagger}\Pi_{y(\bx)}U(\bx)|\psi}=1-\braket{\psi|H_{\bx}|\psi}$. 
A direct calculation shows that the training step (iii) effectively updates the state according to  (Supplementary Sec. II E):
\begin{equation}\label{equ:update rule}
    \ket{\psi}\leftarrow \frac{(\mathbf{I}-\eta H_{\bx})\ket{\psi}}{\norm{(\mathbf{I}-\eta H_{\bx})\ket{\psi}}},
\end{equation} 
where $\norm{(\mathbf{I}-\eta H_{\bx})\ket{\psi}}$ is a normalization factor whose square gives the success probability of post-selection. We define a loss function as the average failure probability to predict the label of a random training datum: $\hat{R}_S(\psi) = \E_{\bx\sim S}\braket{\psi|H_{\bx}|\psi}=\braket{\psi|H_S|\psi}$,
where $\E_{\bx\sim S}$ denotes the expectation with $\bx\sim S$ meaning $\bx$ is uniformly sampled from the training set $S$ and $H_S=\E_{\bx\sim S} H_{\bx}$ is the averaged Hamiltonian. The loss defined in this way has a clear physical meaning: it is just the energy of $\ket{\psi}$ under the Hamiltonian $H_S$. As a result, finding its global minimum is equivalent to finding the ground state energy of $H_S$.

For convenience, we rewrite Eq.~\eqref{equ:update rule} in the density matrix formalism without normalization: $\rho\leftarrow (\mathbf{I}-\eta H_{\bx})\rho (\mathbf{I}-\eta H_{\bx})$. This way, $\Tr(\rho)$ is the success probability of the post-selection and the density matrix keeps track of the overall success probability. Another benefit of the density matrix formalism is that we can embed the randomness of the sample into the density state. Noting that $\bx$ is uniformly sampled from $S$, the averaged post-selection state after each training step (iii) reads $\rho \leftarrow e^{-\eta H_S}\rho e^{-\eta H_S} + \mathcal{O}(\eta^2)$ (Methods). Denoting the initial state as $\rho_0$, the averaged state after $T$ training steps is
\begin{equation}\label{Eq:density Tsteps}
    \rho = e^{-\eta T H_S}\rho_0 e^{-\eta T H_S} + \mathcal{O}(T\eta^2).
\end{equation}
From Eq.~\eqref{Eq:density Tsteps}, the dynamics of $\rho$ is determined by $\beta=\eta T$, the summation of learning rates in all training steps. By choosing $\eta$ sufficiently small, the higher-order terms diminish and Eq.~\eqref{Eq:density Tsteps} reduces to $\rho_{\beta} = e^{-\eta T H_S}\rho_0 e^{-\eta T H_S}$, which is exactly the imaginary-time evolution of $\rho_0$ under the Hamiltonian $H_S$. We remark that for a given machine learning task, the size of the input data sample is typically finite. As a result, $H_S$ would maintain a finite energy gap between the ground and excited states. Consequently, the training process will converge exponentially towards the ground state as $T$ increases until the higher-order terms become significant. We summarize the convergence analysis in the following theorem  (Supplementary Sec. III B):
\begin{theorem}\label{Thm1}
    Suppose $\rho_0$ has a nonzero overlap with the ground space of $H_S$. For an arbitrarily small constant $c$, we can choose an appropriate $\eta$ and $T$ such that the QAL protocol converges exponentially towards the global minimum up to higher-order corrections and the final averaged loss is upper bounded by $E_g+c$, where $E_g$ denotes the ground state energy of $H_S$.
\end{theorem}

Theorem~\ref{Thm1} implies that the convergence of the QAL protocol is guaranteed, as long as the initial state $\rho_0$ has a nonzero overlap with the ground space of $H_S$, which in turn can be assured by initializing $\rho_0$ to be the maximally-mixed state $\rho_0=\mathbf{I}/2^n$. In practice, a random state would have a nonzero overlap with the ground space of $H_S$ with almost unity probability and hence a random initial state $\rho_0$ suffices. We mention that the QAL training process can also be understood from a more familiar variational perspective in machine learning: it indeed implements gradient descent in an automated fashion (Supplementary Sec. II E). In fact, one may regard $\ket{\psi}$ as a variational state parametrized by a complex vector $\psi$. For each single datum, the gradient of $\braket{\psi|H_{\bx}|\psi}$ with respect to $\psi$ is $2H_{\bx}\ket{\psi}$. Therefore, the update rule  in Eq.~\eqref{equ:update rule} essentially implements the stochastic projected gradient descent algorithm to minimize the loss function $\hat{R}_S(\psi)$ with a batch size one. 
From this perspective, our QAL approach essentially exchanges the roles of data and variational parameters.  In conventional quantum neural networks, the input data are encoded into quantum states and the variational parameters specify the quantum circuits. Whereas, in the QAL approach we use data to specify the quantum circuits and treat quantum states as variational parameters. This is reminiscent of the shift from the Heisenberg picture to the Schr\"{o}dinger picture to some extent~\cite{Griffiths2019Introduction}: the conventional approach resembles the Heisenberg picture, where the ``operators'' (variational quantum circuits) are updated during the training process; in contrast, the QAL approach is more akin to the  Schr\"{o}dinger picture, where one evolves the ``states'', rather than the variational ``operators'',  to train the model.

\vspace{.5cm}
\noindent \textbf{\large{}A subtle trade-off}{\large\par}

\noindent In the QAL protocol, we exploit post-selection to implement the desired data-dependent dissipation during the training process. This gives rise to a concern that the overall success probability may decay exponentially with the number of steps. In this section, we show that there is a subtle trade-off between the prediction accuracy and the overall post-selection success probability:  we prove that, as long as $\rho_0$ has a constant overlap with the low energy eigenspace of $H_S$, by choosing an appropriate $\beta$, we can achieve a near-optimal training loss with a constant success probability. We have the following theorem as proved in Supplementary Sec. III D:

\begin{theorem}\label{thm:analytical}
    Assume the spectrum of $H_S$ has a heavy tail, that is, has a constant proportion of low-energy eigenstates. With a random initial state $\rho_0$ in the computational basis, an appropriate learning rate $\eta$, and an appropriate number of steps $T$, the final state $\rho_{\beta}$ achieves a near-optimal training loss with a constant success probability.
\end{theorem}

\begin{figure*}
    \centering
    \includegraphics[width=\linewidth]{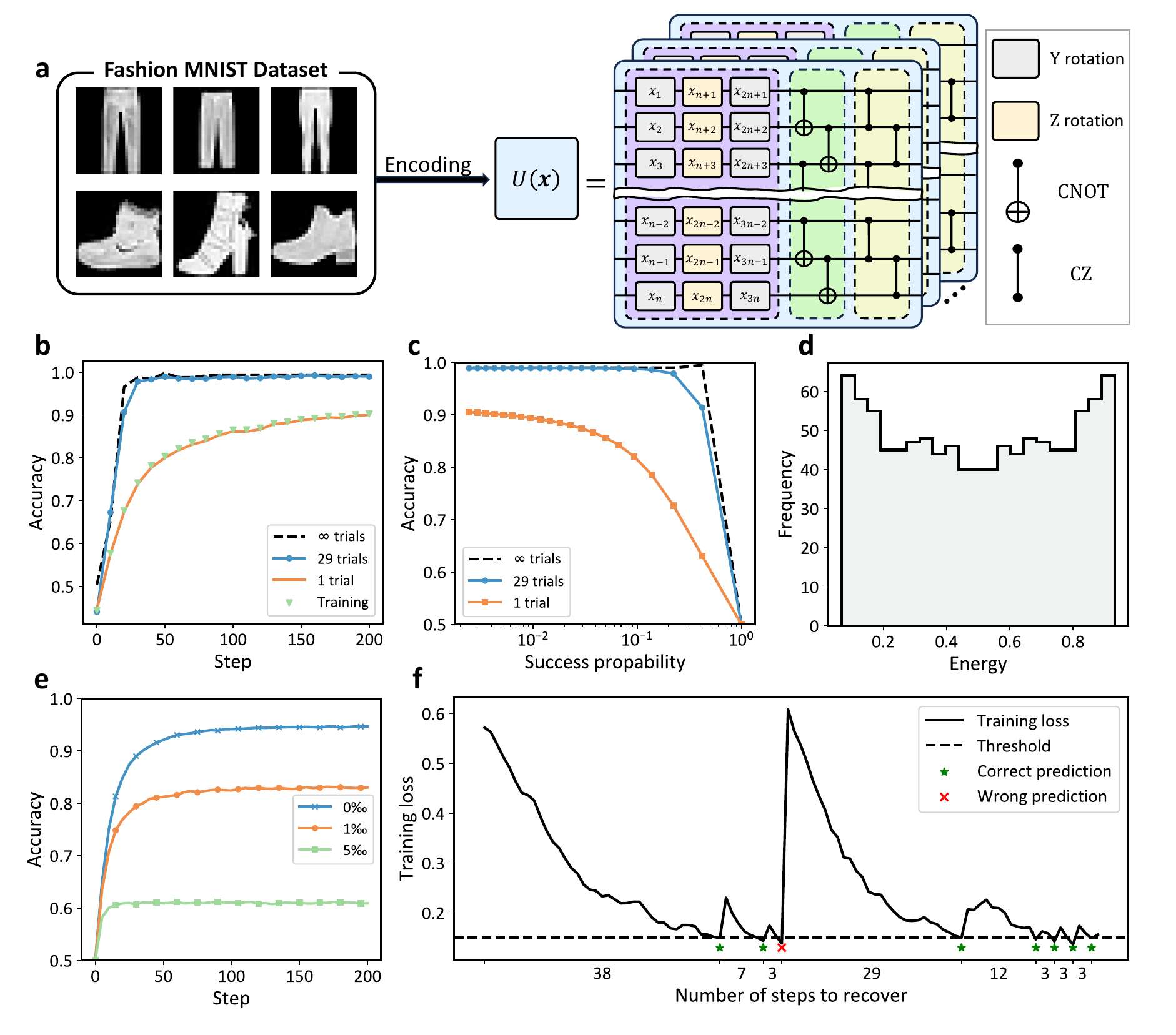}
    \caption{
    \textbf{Classify images in the Fashion MNIST dataset~\cite{xiaoFashionMNISTNovelImage2017} using QAL.} \textbf{a}, Example images from the dataset belonging to the classes ``trouser" and ``ankle boot". The pixel values of an image are encoded into the rotation angles of variational single-qubit gates in a quantum circuit. Here, CNOT and CZ denote the controlled-NOT and controlled-Z gates, respectively. \textbf{b}, Testing and training accuracy during the training process. We also plot the accuracy amplified by majority vote with multiple trials, see Supplementary Sec.~II D for the details of the majority vote. \textbf{c}, The trade-off between testing accuracy and the success probability of post-selection. The success probability is analytically calculated from Eq. \eqref{Eq:density Tsteps}, where we omit the higher-order term $O(T\eta^2)$. \textbf{d}, Spectrum of the Hamiltonian $H_S$ associated with the training dataset. \textbf{e}, Training performance under depolarizing noise with different two-qubit gate noise strengths. The single-qubit noise rate is set to one-tenth of the two-qubit noise rate. The green line (5\textperthousand) corresponds to the noise level of real-world superconducting quantum devices~\cite{jinObservationTopologicalPrethermal2025}. \textbf{f}, Illustration of state reusability: whenever the training loss drops below a threshold of 0.15, a prediction is made while training continues. The x-axis represents the number of steps required to recover the performance.} 
    \label{fig:numerics}
\end{figure*}

We remark that the heavy-tail assumption is reasonable in machine learning scenarios---it stems from the fact that data samples with the same label should bear similar data structures and thus correspond to similar Hamiltonians. Consider a simple example of classifying images of dogs and cats, where all dogs look similar and all cats look similar. The Hamiltonian $H_S$ is approximately a mixture of two projectors of dimensions $2^{n-1}$, $H_{\text{dogs}}$ and $H_{\text{cats}}$. Regarding $H_{\text{dogs}}$ and $H_{\text{cats}}$ as random projectors, then $H_S$ has a constant proportion of near-zero eigenvalues, thus has a heavy tail. The heavy-tail assumption is further verified with real-life datasets, as shown in Fig.~\ref{fig:numerics}\textbf{d}. We note that heavy-tail assumption does not hold for typical Hamiltonians encountered in quantum physics, which in general exhibit exponential concentration of energy levels around the middle spectra.

\vspace{.5cm}
\noindent \textbf{\large{}Bounded generalization error}{\large\par}
\noindent 
The above discussions have established a provable and explainable trainability for the QAL protocol. Yet, its generalization ability remains uncharted: it is unclear whether the good performance on the training dataset can be generalized to unseen data samples. We address this crucial issue in this section by proving an upper bound for the generalization error. Defining the true loss function as $R(\psi)=\E_{\bx\sim \cD}\braket{\psi|H_{\bx}|\psi}$, the following theorem bounds the generalization gap with the rigorous proof given in Supplementary Sec. III E:
\begin{theorem}\label{thm:generalization}
    Supposing the data samples are drawn randomly and independently from $\cD$, then with probability at least $1-\delta$, the generalization gap is upper bounded by
    \begin{equation}
        \max_{\psi}\big(R(\psi)-\hat{R}_S(\psi)\big) \leq \sqrt{\frac{4\ln(2^{n+1}/\delta)}{N}}.
    \end{equation}
\end{theorem}

According to this theorem, as long as the size $N$ of the training dataset is larger than $\Omega(n)$ (i.e., the logarithm of the degree of freedom), a  state $\ket{\psi}$ with low training loss has a low generalization loss with high probability. 
The good generalization of our QAL approach stems from the simple quadratic form of the loss function.

\vspace{.5cm}
\noindent \textbf{\large{}State reusability}{\large\par}
\noindent
In classical machine learning or conventional gradient-based quantum learning models, variational parameters are usually stored in a classical computer and they can be copied and reused on demand during the inference stage. In contrast, in the QAL approach what we obtain after training is a particular quantum state that carries information about the training samples. Due to the quantum no-cloning theorem~\cite{Wootters1982Single}, this state cannot be copied. This, together with the fact that the quantum measurements involved are destructive, results in a possible drawback of the QAL approach: it seems that one needs to carry out repeatedly the whole training process to prepare the desired state for each run of the inference. Fortunately, this is not the case.

There are two complementary ways to solve this problem. First, we can incorporate quantum shadow tomography techniques ~\cite{Aaronson2018Shadow}  into step (v) of the QAL protocol. This will substantially reduce the number of copies of the desired states during inference: $\text{polylog}(N_p)$ copies would be sufficient for classifying $N_p$ unseen data samples. Second, one may reuse the state after the measurements in step (v). Theoretically, the state after training $\ket{\psi^*}$ would output correct labels for unseen sample $\bx$ with high probability, as proved above that the QAL process converges to the global minimum with bounded generalization error. Hence, the measurement $\{U(\bx)^\dagger \Pi_{y(\bx)} U(\bx)\}_y$ is in fact a gentle measurement that would not disturb $\ket{\psi^*}$ too much. Although the post-measurement state is different from $\ket{\psi^*}$, a few more training steps would recover the system to a state with high prediction accuracy. This reusability of states is verified by extensive numerical simulations on different real-life datasets (Fig.~\ref{fig:numerics}\textbf{f}). One can combine shadow tomography with state reusability to further reduce the resources required during the inference step.

\vspace{.5cm}
\noindent \textbf{\large{}Numerical results}{\large\par}
\noindent To further illustrate and benchmark the effectiveness of QAL, we carry out extensive numerical simulations on a range of datasets, including real-life images (e.g., hand-writing digit images) and quantum data (e.g., thermal and localized quantum many-body states). In Fig.~\ref{fig:numerics}, we plot our numerical results for classifying images in the Fashion MNIST dataset~\cite{xiaoFashionMNISTNovelImage2017}. Fig.~\ref{fig:numerics}\textbf{a} shows some samples from the dataset and the encoding scheme used.  Fig.~\ref{fig:numerics}\textbf{b} plots the training and testing accuracy versus training iterations, from which it is clear that the accuracy increases rapidly and then saturates at a high value of $0.99$. In addition, we find that this approach is surprisingly efficient in the sense that one only need to run the experiment very few times to achieve a relatively high accuracy. From Fig.~\ref{fig:numerics}\textbf{b}, the accuracy for $29$ trials is already very close to that for infinite trials. In Fig.~\ref{fig:numerics}\textbf{c}, we plot the training accuracy versus the overall post-selection success probability. From this figure, the accuracy is maintained at a high level up to a sizable success probability. For this particular example, it remains about $0.99$ up to a success probability larger than $0.1$ for only $29$ trials.  Therefore, to achieve an accuracy as high as $0.99$, we only need to run the training circuit for roughly $290$ times, which can be done in less than a  second on a superconducting quantum device~\cite{Kjaergaard2020Superconducting}. As a comparison, for conventional gradient-based approaches, $290$ executions of the variational circuit is barely enough to estimate the gradient of a single parameter in one training step.

In Fig.~\ref{fig:numerics}\textbf{d}, we plot the spectrum of the averaged Hamiltonian $H_S$. From this figure, it is evident that $H_S$ indeed has a large portion of low-energy eigenstates, validating the heavy-tail assumption. Another salient feature of the QAL protocol is its robustness to experimental noise, which has an origin in the dissipation nature in preparing the target state $\ket{\psi^*}$ during training. This has been verified in our numerical simulations, as shown in Fig.~\ref{fig:numerics}\textbf{e}. In Fig.~\ref{fig:numerics}\textbf{f}, we show the state reusability for the QAL protocol. We keep training the model and predict an unseen datum whenever the training loss is below a threshold value of $0.15$. Most predictions are correct, in which case the performance for the post-measurement state is not degraded too much (as discussed above) and is recovered after about three additional training steps. When the prediction is wrong, its performance can be even worse than that for the random initial state. Nevertheless, the training steps needed to recover can be notably smaller, indicating that partial of the data information is preserved in the post-measurement state even if the measurements give wrong predictions. Our numerical simulations for classifying images of hand-written digits in the MNIST dataset and symmetry-protected topological states are plotted in Supplementary Sec. IV, where similar results are obtained.

\vspace{.5cm}

{\noindent \textbf{\large{}Discussion and outlook}{\large\par}}

\noindent 
The introduced QAL protocol is gradient-free and thus escapes the barren plateau problem inherently. It features a number of striking merits such as provable and explainable trainability with bounded generalization error.  Yet, several questions of fundamental importance remain unsolved. First, in conventional gradient-based approaches, quantum neural networks are shown to possess universal representation power. They can approximate an arbitrary function to arbitrary accuracy as long as the number of variational parameters is large enough~\cite{Goto2021Universal}. In contrast, the QAL protocol involves no variational parameter and its representation power depends on particular encoding schemes. Intuitively, one can always exploit different encoding schemes to approximate an arbitrary function. However, a rigorous proof of the universal representation power for the QAL protocol is technically challenging and remains unknown.  Second, the demonstration of quantum advantages is a long-sought-after goal in the field of quantum machine learning. It would be interesting and important to prove in theory and demonstrate in experimental quantum advantages for certain learning tasks in the QAL framework. Third, our discussions mainly focus on supervised learning. An extension of the QAL protocol to unsupervised and reinforcement learning scenarios is well worth exploring.

Given the fact that the QAL protocol bears a certain degree of robustness to noise and does not need a quantum random access memory~\cite{Giovannetti2008Quantum} to transfer classical data into quantum states, an experimental demonstration of such a protocol with current noisy intermediate-scale quantum devices is highly feasible and desirable. Such an experiment would be a crucial step toward large-scale practical applications of quantum technologies in artificial intelligence.

\vspace{.5cm}
\noindent\textbf{\large{}Methods}{\large\par}

\vspace{.2cm}
\noindent\textbf{Implementation of the  target-oriented perturbation}{\large\par}
\noindent
During the training process, the state $\ket{\psi}$ is updated in step (iii) by a non-unitary perturbation $M_y=\ketbra{y}+(1-\eta)(\mathbf{I}-\ketbra{y})$. Such a perturbation has a clear physical meaning of suppressing (enhancing) the probability of incorrect (correct) prediction, hence evolving the state $\ket{\psi}$ towards the target state  $\ket{\psi^*}$. This is similar to the Hamiltonian-echo-backprogagation based self-learning approach~\cite{lopez-pastorSelfLearningMachinesBased2023} for classical learning, where a small error signal is injected on top of the evaluation field to guide the training process. To implement $M_y$, we consider adding an ancillary qubit and embedding $M_y$ into a unitary $U_y$: 
\begin{equation}
    U_y = M_y\otimes Z + \sqrt{I-M_y^2}\otimes X,
\end{equation}
where $Z$ and $X$ are the Pauli matrices acting on the ancillary qubit. We initialize the ancillary qubit to state $\ket{0}$ and then apply $U_y$ to $\ket{\psi}\otimes\ket{0}$. Noting that $(\mathbf{I}\otimes \bra{0})U_y(\mathbf{I}\otimes \ket{0}) = M_y$, the state $\ket{\psi}$ will be updated by $M_y$ after we measure the ancillary qubit in the computational basis and post-select the outcome $0$. In this way, we effectively implement the non-unitary perturbation $M_y$ by adding an ancillary qubit combined with post-selection. 

It is worthwhile to remark that $U_y$ can be implemented with elementary gates very efficiently. For a $k$-class classification task, $M_y$ acts on $\lceil{\log k}\rceil$ qubits, where $\lceil\cdot\rceil$ denotes the ceiling function that outputs the least integer greater than or equal to the input. $U_y$ acts on $\lceil{\log k}\rceil+1$ qubits and is a multi-controlled gate up to single-qubit gates. It can be efficiently compiled into elementary gates with gate complexity scaling linearly in $\log k$ (See Supplementary Sec II C). We note that such a gate complexity scaling is much better than that for compiling a general unitary with the Solovay-Kitaev algorithm~\cite{Dawson2006TheSolovay}, where a $2^{O(\log k)}=\text{poly}(k)$ gate complexity is required for a given accuracy.

\vspace{.5cm}
\noindent\textbf{Data encoding}{\large\par}

\noindent For classical classification, the data sample can be represented as a vector $\bx$ (for example, the pixel information of an image). We encode $\bx$ into a quantum circuit consisting of $\lceil{\frac{l}{3n}}\rceil$ blocks with similar structure, where $l$ denotes the dimension of $\bx$. Each block contains a layer of single-qubit rotations parametrized by components of $\bx$ and a layer of two-qubit entangling gates (see Fig.~\ref{fig:numerics}\textbf{a} and Supplementary Sec. II B). Such an encoding scheme features an intriguing merit of bypassing the use of a quantum random access memory~\cite{Giovannetti2008Quantum}, which is resource-demanding and inaccessible so far, to load classical data. This makes the QAL approach attainable for middle-size data samples ($l$ on the order of tens of thousands) with current noisy intermediate-scale quantum devices. For even larger data samples, such as the largest images in the ImageNet dataset~\cite{Deng2009ImageNet} ($l\sim 10^8$), this encoding scheme would become impractical for current quantum devices since the circuit depth is formidably large.  In fact, the required circuit depth scales roughly as $\sim l/(3n)$. In practice, the number of qubits $n$ can be chosen on the order of $\log l$ and hence the circuit depth $\sim l/(3\log l)$. For an image with $l\sim 10^8$, we need a quantum circuit with depth $\sim 10^6$ to encode this image. This is beyond the reach of current noisy quantum devices with state-of-the-art performance~\cite{Xiang2024Longlived, Jin2025Observation, Acharya2408Quantum,Bluvstein2024Logical,Iqbal2024non,Cao2023Generation}. One may use more qubits to encode $\bx$ and reduce the circuit depth.  

For the classification of Hamiltonian data, the encoding is straightforward. One can encode $H_{\bx}$ into real-time evolution $e^{-iH_{\bx}t}$, which in turn can be implemented via quantum simulation techniques~\cite{Childs2012Hamiltonian,Georgescu2014Quantum,Low2017Optimal,Clinton2021Hamiltonian}. In quantum state classification, each datum is a quantum state $\ket{\bx}$ of $s$ qubits. Fix an $(s+n)$-qubit Hamiltonian $H$, $H_{\ket{\bx}}=(\bra{\bx}\otimes \mathbf{I}_n)H(\ket{\bx}\otimes \mathbf{I}_n)$ is an $n$-qubit Hamiltonian. We then encode the state $\ket{\bx}$ into a unitary $U_{\ket{\bx}}=e^{-iH_{\ket{\bx}}}$. This unitary can be efficiently implemented using copies of $\ket{\bx}$ and real-time evolution $e^{-iHt}$, inspired by the Lloyd-Mohseni-Rebentrost protocol~\cite{Lloyd2014Quantum}. See the Supplementary Sec. II B for details.

\vspace{.3cm}
\noindent\textbf{Numerical setup}\\
\noindent We conduct numerical simulations on various types of datasets, including classical data, Hamiltonian data, and quantum state data.
For classical data, we use the Fashion MNIST dataset~\cite{xiaoFashionMNISTNovelImage2017} and focus on the binary classification of classes ``trouser" and ``ankle boot". To reduce the computation overhead, we rescale the figures into $10\times 10$ pixels and then normalize the pixel values. The pixel values are encoded into rotation angles of single-qubit gates in a parameterized quantum circuit. We also conduct QAL protocol to classify MNIST dataset~\cite{lecun1998mnist} and present the results in the Supplementary Information.
For Hamiltonian data, we consider a binary classification of the following Aubry-Andr\'{e} Hamiltonian on $10$ qubits~\cite{aubry1980analyticity}:
\begin{equation*}
H=-\frac g2 \sum_k (\sigma_k^x\sigma_{k+1}^x+\sigma_k^y\sigma_{k+1}^y)-\frac V2 \sum_k \cos(2\pi \phi k)\sigma_k^z, 
\end{equation*}
where $g$ is the coupling strength, $\sigma_k^\alpha$ ($\alpha=x, y, z$) are Pauli operators on the $k$-th qubit, $V$ is the disorder magnitude and $\phi=(\sqrt{5}-1)/2$. This Hamiltonian exhibits a quantum phase transition at $V/g=2$, between a localized phase for $V/g>2$ and a delocalized phase for $V/g<2$. To generate the dataset, we fix $g=1$, sample $V$ in the interval $[0, 4]$ and label the Hamiltonian according to its phase. To carry out the QAL protocol, we encode the Hamiltonian into its real-time evolution $e^{-2iH}$. Our numerical results for Hamiltonian data are ploted in Fig.~S2 in the Supplementary Information. 
For quantum state data, we classify the ground state of $10$-qubit clustering-Ising model Hamiltonian~\cite{son2011quantum,smacchia2011statistical} with periodic boundary condition:
\begin{equation*}
    H(h)=-\sum_{k}\sigma_k^x\sigma_{k+1}^z\sigma_{k+2}^x+h\sum_{k}\sigma_k^y\sigma_{k+1}^y,
\end{equation*}
This Hamiltonian has a phase transition at $h=1$, between a symmetry protected topological phase ($h<1$) and an antiferromagnetic phase ($h>1$). We sample $h$ from interval $[0, 2]$ to generate the dataset. The detailed data encoding scheme and the numerical results are shown in Supplimentary Sec. IV.

In most numerical simulations, we uniformly set the learning rate to 0.1 and examine the performance on a test dataset of size 500. The only exception is the illustration of noise robustness shown in Fig.~\ref{fig:numerics}\textbf{e}, where to reduce the computational overhead of density matrix simulation, we further thumbnail the Fashion MNIST image to $5\times 5$, increase the learning rate to $0.2$ and perform the simulation with five qubits. 

\vspace{.6cm}
\noindent\textbf{\large{}Data availability}
All the data for this study will be  made publicly available for download on Zenodo/Figshare/Github upon publication.

\vspace{.6cm}
\noindent\textbf{\large{}Code availability}
The data analysis and numerical simulation codes for this study will be made publicly available for download on Zenodo/Figshare/Github upon publication.

\vspace{.5cm}
\noindent\textbf{Acknowledgement} We thank Andrew Chi-Chih Yao, Adi Shamir, Xun Gao, Sirui Lu, Zidu Liu, and Fangjun Hu for helpful discussions.  This work is supported by the National Natural Science Foundation of China (Grant No.~T2225008, No.~T24B2002, and No.~12075128), the Innovation Program for Quantum Science and Technology (Grant No.~2021ZD0302203, 2021ZD0301601, and 2021ZD0301605), the Ministry of Science and Technology of China (2021AAA0150000), Tsinghua University Initiative Scientific Research Program, and the Ministry of Education of China, the Tsinghua University Dushi Program, and the Shanghai Qi Zhi Institute Innovation Program SQZ202318 and SQZ202317. L.-M.D. acknowledges in addition support from the New Cornerstone Science Foundation through the New Cornerstone Investigator Program.

\end{document}


\title{Supplementary Information for: \\Quantum automated learning with provable and explainable trainability}

\maketitle
\tableofcontents

\makeatletter
\setcounter{figure}{0}
\setcounter{equation}{0}
\setcounter{table}{0}
\setcounter{definition}{0}
\setcounter{theorem}{0}
\setcounter{lemma}{0}
\setcounter{proposition}{0}
\renewcommand{\thefigure}{S\@arabic\c@figure}
\renewcommand{\thetable}{S\@arabic\c@table}
\renewcommand{\theequation}{S\@arabic\c@equation}
\renewcommand{\thedefinition}{S\@arabic\c@definition}
\renewcommand{\thetheorem}{S\@arabic\c@theorem}
\renewcommand{\thelemma}{S\@arabic\c@lemma}

\section{Theoretical Background}
\subsection{Supervised learning}
We start by introducing the basic framework of supervised learning \cite{Goodfellow2016Deep}. Let $\cX$ be the set of input data and $\cY=\{1, 2, \cdots, k\}$ be the set of labels. We assume that every input data $\bx\in\cX$ has a deterministic label $y(\bx)\in \cY$. Let $\cD$ be an unknown distribution over $\cX$. The goal of supervised learning is to find an algorithm $\cA(\cdot)$ (probably randomized in quantum machine learning) such that, input a sample $\bx\sim \cD$, output the label $y(\bx)$ with high probability. To achieve this goal, we parametrize the learning model by parameters $\btheta$ and optimize the average loss
\begin{equation}
    R(\btheta)= \E_{\bx\sim \cD} L(\bx, y(\bx); \btheta).
\end{equation}
Here $L(\bx, y; \btheta)$ is some loss function, usually a metric of the difference between the output distribution of $\cA(\bx; \btheta)$ and the correct label $y$. $R(\btheta)$ is called the risk or the prediction error of the model $\cA(\cdot~; \btheta)$. However, the distribution $\cD$ is unknown, so we cannot directly calculate $R(\btheta)$. Instead, we sample a training dataset $S=\{(\bx_i, y_i=f(\bx_i))\}_{i=1}^m$ from $\cD$, and optimize the following empirical risk or training error:
\begin{equation}
    \hat{R}_S(\btheta) = \frac{1}{m}\sum_{i=1}^m L(\bx_i, y_i; \btheta).
\end{equation}
According to the simple decomposition $R(\btheta)=\hat{R}_S(\btheta)+(R(\btheta)-\hat{R}_S(\btheta))$, the success of supervised learning depends on two important factors: trainability and generalization. In short, trainability asks whether we can efficiently find $\btheta$ with low empirical risk, while generalization asks whether the generalization gap $\mathrm{gen}_S(\btheta)=R(\btheta)-\hat{R}_S(\btheta)$ is upper bounded, i.e., whether the good performance on the training set $S$ can be generalized to unseen data.

\subsection{Variational quantum learning models}

For conventional gradient-based quantum learning approaches~\cite{Cerezo2021Variational}, a learning algorithm $\cA(\bx;\btheta)$ executes a variational quantum circuit $U(\btheta)$ to a data-encoded state $\ket{\phi(\bx)}$ before performing certain measurements to make the prediction.  
Assuming the measurement observable to be $O_M$, the output from the variational circuit is the expectation value $\braket{\phi(\bx)|U(\btheta)^\dagger O_MU(\btheta)|\phi(\bx)}$.
The loss function is often defined as a function of this value, where commonly used forms include mean square error and cross-entropy.
For a training task, the average loss value over a given set of training data is defined as the empirical risk,
where schemes based on gradient descents are widely exploited to minimize it and find the optimal parameters $\btheta^*$. 
In the quantum machine learning realm, there are various methods proposed to calculate the gradients with respect to circuit parameters, including finite differences, the parameter-shift rules, and quantum natural gradients~\cite{Mitarai2018Quantum,Schuld2019Evaluating,Stokes2020Quantum}.

Quantum neural networks have demonstrated promising generalization capabilities in various learning settings~\cite{caiSampleComplexityLearning2022,caroGeneralizationQuantumMachine2022}.
Intuitively, when the number of training data points exceeds the degrees of freedom in the parameter space, the generalization gap of the optimized parameters is typically bounded by a small constant.
However, the practical trainability of quantum neural networks remains a significant challenge.
A key bottleneck lies in the computational cost of estimating gradients with respect to the circuit parameters. 
For instance, computing the gradient of a single parameter accurately often requires executing the variational circuit thousands of times, even when employing comparably efficient parameter-shift rules.
This process becomes increasingly time-consuming and impractical as the number of parameters grows.

Furthermore, the loss landscape of quantum neural networks can be highly non-convex and challenging to navigate.
As shown in ref.~\cite{youExponentiallyManyLocal2021}, the loss function of quantum neural networks exhibits exponentially many local minima, which can trap optimization algorithms and hinder convergence.
In parallel, the phenomenon of barren plateaus, first identified in ref.~\cite{McClean2018Barren}, poses a critical issue: the gradients of the loss function tend to vanish exponentially with the number of qubits, especially in deep quantum circuits. In such cases, the loss landscape becomes effectively flat, making it extremely difficult to identify a direction for optimization. The presence of barren plateaus is closely tied to the randomness and entanglement structure of the quantum circuit, as well as the choice of cost function and initial parameterization, which severely threads the scalability and practical utility of quantum neural networks for large-scale problems~\cite{McClean2018Barren,Cerezo2021Cost,OrtizMarrero2021EntanglementInduced,Holmes2022Connecting,Wang2021Noiseinduced,Larocca2024Review}.

\section{The Automated Learning Strategy}
In this section, we provide more technical details about the quantum automated learning strategy.

\subsection{Choose an appropriate number of qubits}
To carry out the QAL protocol, the first step is to decide an appropriate number of qubits $n$. Since an $n$-qubit state lives in a $O(2^n)$-dimensional Hilbert space and thus bears $O(2^n)$ degrees of freedom, one natural choice is $n=O(\log\abs{\bx})$, where $\abs{\bx}$ is the dimension of data sample $\bx$. However, we remark that the choice of $n$ is much more flexible. For example, if we are classifying Hamiltonian data, it is more natural to set $n$ to be the system size of the Hamiltonian. If the data are  images of size $L\times L$, setting $n=L$ may align with the two-dimensional structure better. On the other hand, sometimes it is possible to set $n$ to be even much smaller than $\log(\abs{\bx})$,  since realistic data samples are usually believed to lie in a low-dimension manifold. 

\subsection{Encode data into unitaries}\label{supp_sec:data_encoding}
\begin{figure}[htbp]
    \centering
    \includegraphics[width=0.9\linewidth]{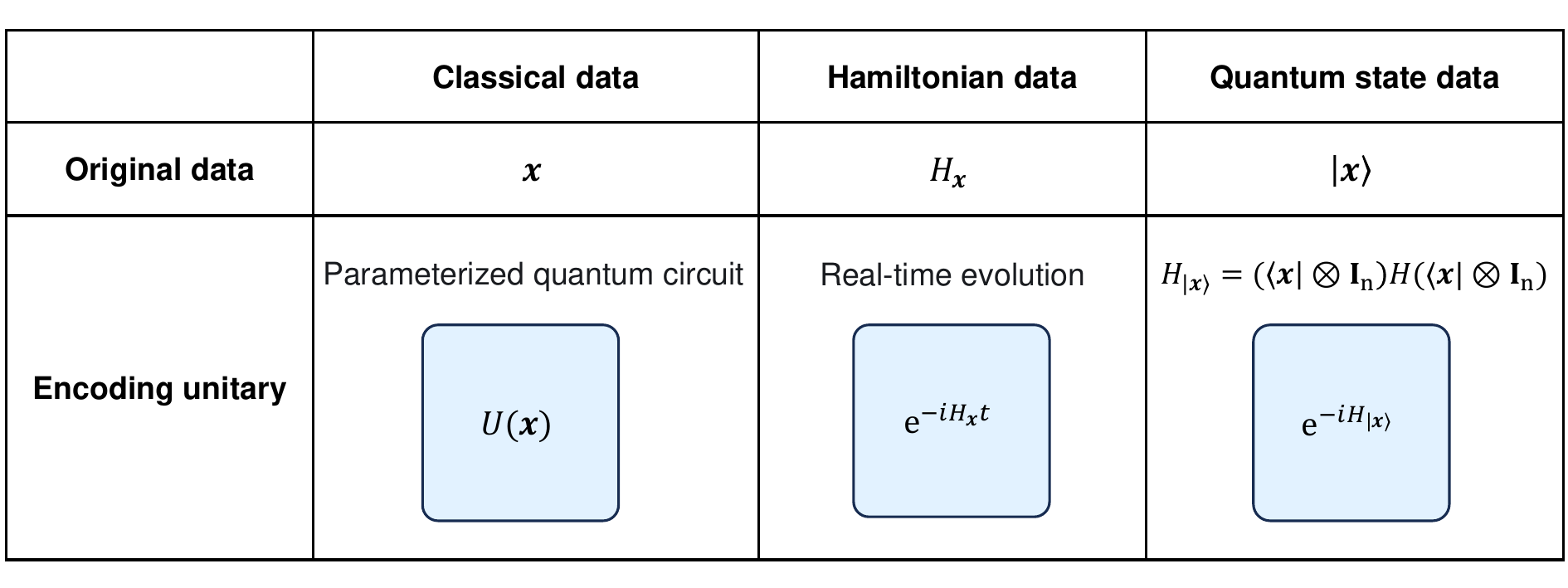}
    \captionsetup{justification=raggedright, singlelinecheck=false}
    \caption{\textbf{Overview of three different data encoding methods.} We encode classical data $\bx$ into parameterized quantum circuit $U(\bx)$; encode Hamiltonian data $H_{\bx}$ into its real-time evolution $e^{-iH_{\bx}t}$; and encode quantum state $|\bx\rangle$ first into an $n$-qubit Hamiltonian: $H_{\ket{\bx}}=(\bra{\bx}\otimes \mathbf{I}_n)H(\ket{\bx}\otimes \mathbf{I}_n)$, and then encode it to a unitary $e^{-iH_{\ket{\bx}}}$.}
    \label{fig:3data_encode}
\end{figure}

Once we pin down the number of qubits $n$, the next step is to encode data into $n$-qubit unitaries. Here we present the detailed data encoding schemes for quantum automated learning, which incorporates three distinct categories of data: 
classical data, Hamiltonian data, and quantum state data.
An overview of the encoding methods is provided in Fig.~\ref{fig:3data_encode}, which summarizes the key approaches before delving into the detailed descriptions of each scheme. 

\begin{figure}
    \centering
    \includegraphics[width=\linewidth]{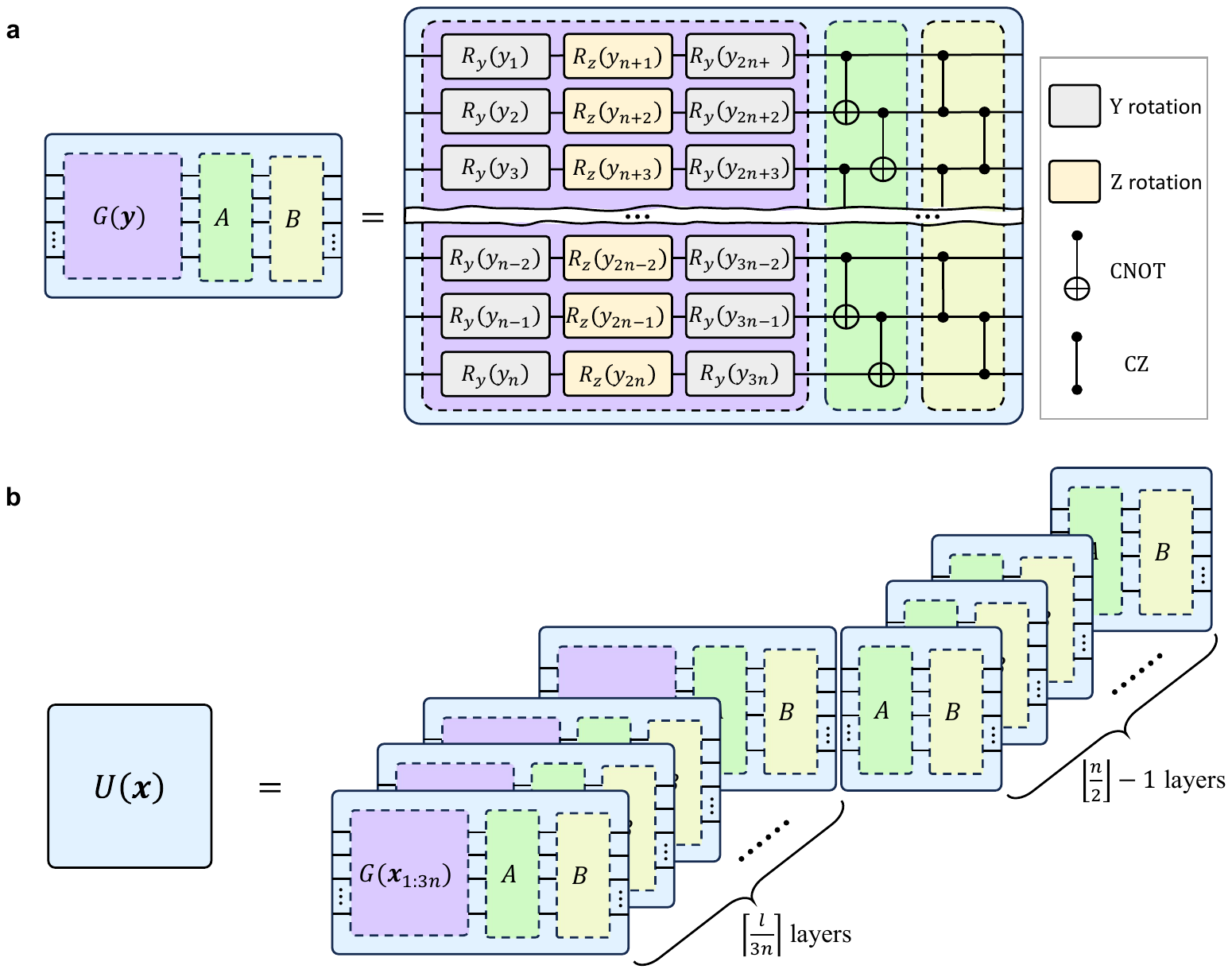}
    \captionsetup{justification=raggedright, singlelinecheck=false}
    \caption{\textbf{The classical data encoding scheme.} \textbf{a}, Illustrates an encoding layer for a $3n$-dimensional vector $\textbf{y}$ ($G\left(\textbf{y}\right)$). The layer consists of three layers of single-qubit gates ($R_y, R_z, R_y$), denoted by the purple block; two layers of CNOT two-qubit gates, denoted by the green block; and two layers of CZ two-qubit gates, denoted by the yellow block. \textbf{b}, Illustrates the complete encoding for an $l$-dimensional vector $\bx$. The encoding scheme involves $\lceil \frac{l}{3n}\rceil$ layers of single-qubit and two-qubit gates ($BAG\left(\bx_{3n(k-1)+1:3nk}\right)$), followed by $\lfloor \frac n 2 \rfloor -1$ layers of entangling gates ($BA$).}
    \label{fig:data_encode}
\end{figure}

Classical data, including images, text, or audio, can be transformed into a vector of numerical values, denoted as $\bx$. The data vector $\bx$ is encoded into parameterized quantum circuit. Specifically, each element of $\bx$ is mapped into rotation angles of single-qubit gates. A single-qubit gate is parametrized as $G(\alpha, \beta, \gamma)=R_y(\alpha)R_z(\beta)R_y(\gamma)$, where $R_y(\alpha)$ and $R_z(\beta)$ are the rotations around the $Y$ and $Z$ axes of the Bloch sphere by angle $\alpha$ and $\beta$, respectively. Therefore, for a n-qubit quantum circuit, a layer of single-qubit gates can encode up to $3n$ entries of the vector $\bx$. If we denote the dimension of $\bx$ as $l$, then it is necessary to employ $\lceil \frac{l}{3n}\rceil$ layers of single-qubit gates. More concretely, considering a $3n$-dimensional vector $\textbf{y}$, we define the encoding of a single-qubit layer as:
$G(\textbf{y})=\otimes_{i=1}^n G_i(y_{2n+i}, y_{n+i}, y_{i})$, 
where $G_i$ acts on the $i$-th qubit, as illustrated in Fig.~\ref{fig:data_encode}\textbf{a}. Then the $k$-th layer single-qubit encoding of the    data vector $\bx$ is defined as $G\left(\bx_{3n(k-1)+1:3nk}\right)$, where $\bx_{i:j}$ denotes the abbreviation of $(x_i,x_{i+1},\dots,x_j)$. In cases where the number of elements in $\bx$ does not exactly divide by $3n$, padding with zeros is used to ensure uniformity.

Between two layers of single-qubit gates, we insert a layer of two-qubit gates to entangle the qubits, leading to the spread of information. This layer of two-qubit gates is composed of a CNOT-gate block $A$ and a CZ-gate block $B$. Each block consists of two layers of two-qubit gates: in the first layer, the odd-numbered qubits act as the control qubits, while in the second layer, the even-numbered qubits serve as the control qubits. In both layers, each control qubit targets the subsequent qubit in the sequence. Mathematically, we define: $A=\left(\otimes_{i=1}^{\lfloor \frac {n-1}{2} \rfloor} \text{CNOT}_{2i,2i+1}\right) \left(\otimes_{i=1}^{\lfloor \frac n 2 \rfloor} \text{CNOT}_{2i-1,2i}\right)$ and $B=\left(\otimes_{i=1}^{\lfloor \frac {n-1}{2} \rfloor} \text{CZ}_{2i,2i+1}\right) \left(\otimes_{i=1}^{\lfloor \frac n 2 \rfloor} \text{CZ}_{2i-1,2i}\right)$, as shown in Fig.~\ref{fig:data_encode}\textbf{a}. To ensure that all elements of the data vector can influence the measured qubits used for prediction, we add additional $\lfloor \frac{n}{2} \rfloor -1$ layers of two-qubit gates. 

The final unitary encoding $U(\bx)$ for classical data is then given by a sequence of single- and two-qubit gates:

\begin{equation}
    (BA)^{\lfloor \frac{n}{2} \rfloor -1}G(\bx_{3n(d-1)+1:3nd})\cdots BAG(\bx_{3n+1:6n})BAG(\bx_{1:3n}),
\end{equation}
where $d=\lceil \frac{l}{3n}\rceil$ and $\bx$ is padded with zeros if $3nd>l$.
as illustrated in Fig.~\ref{fig:data_encode}\textbf{b}.

For Hamiltonian data, we encode the Hamiltonian $H_{\bx}$ through its real-time evolution $e^{-iH_{\bx} t}$. This encoding inherently captures the time evolution of quantum states governed by the Schr\"{o}dinger equation. Both $e^{-iH_{\bx} t}$ and its reverse evolution $e^{iH_{\bx} t}$ can be implemented through quantum Hamiltonian simulation techniques \cite{Childs2012Hamiltonian,Georgescu2014Quantum,Low2017Optimal,Clinton2021Hamiltonian}. One may exploit other encoding schemes for $H_{\bx}$. In practice, we find that our real-time evolution encoding works well, as shown in our numerical simulations (Fig. \ref{fig:AA})

For quantum state classification, each datum is a quantum state $\ket{\bx}$ of  $s$ qubits. In order to carry out the QAL protocol for classifying $\ket{\bx}$, we need to encode $\ket{\bx}$ into a unitary $U_{\ket{\bx}}$. 
We fix an $(s+n)$-qubit Hamiltonian $H$.  We first encode the quantum state $|\bx\rangle$ into an $n$-qubit Hamiltonian: $H_{\ket{\bx}}=(\bra{\bx}\otimes \mathbf{I}_n)H(\ket{\bx}\otimes \mathbf{I}_n)$, and then encode it to a unitary $U_{\ket{\bx}}=e^{-iH_{\ket{\bx}}}$. This unitary can be efficiently implemented using copies of $\ket{\bx}$ and real-time evolution $e^{-iHt}$, inspired by the Lloyd-Mohseni-Rebentrost protocol \cite{lloydQuantumPrincipalComponent2014} that implements $e^{-i\rho t}$ from copies of $\rho$. More concretely, for any state $\rho$ straightforward calculations yield
\begin{align}
    \tr_{\leq l}(e^{-iH\Delta t}(\ketbra{\bx}\otimes \rho)e^{iH\Delta t})&=\rho + \tr_{\leq l}(-iH\Delta t (\ketbra{\bx}\otimes \rho)+(\ketbra{\bx}\otimes \rho)iH\Delta t))+O((\Delta t)^2)\nonumber\\
    &= \rho - i\Delta t [H_{\ket{\bx}}, \rho] + O((\Delta t)^2)\nonumber\\
    &= e^{-iH_{\ket{\bx}}\Delta t}\rho e^{iH_{\ket{\bx}}\Delta t} + O((\Delta t)^2).
\end{align}
Therefore, if we apply $e^{-iH\Delta t}$ to $\ketbra{\bx}\otimes \rho$, we effectively apply $e^{-iH_{\ket{\bx}}\Delta t}$ to $\rho$, up to a second-order error $O((\Delta t)^2)$. Repeating this procedure $1/\Delta t$ times, we effectively apply $e^{-iH_{\ket{\bx}}}$ to $\rho$ up to error $O(\Delta t)$. By choosing $\Delta t$ sufficiently small, we can approximate $e^{-iH_{\ket{\bx}}}$ up to any precision. In our numerical simulations, we choose $H$ as a random 4-local Hamiltonian. Each term of $H$ is a 4-body Pauli interaction on 4 random positions with a random interaction strength between -1 and 1 (Fig. \ref{fig:CIM}).

\subsection{Training process and efficient compiling of $U_y$}
In this subsection, we show that the required unitary $U_y$ to implement the target-oriented perturbation can be compiled in an efficient way, with the count of CNOT gates scales logarithmically with the number of classes $k$. Recalling that $U_y=M_y\otimes Z + \sqrt{I-M_y^2}\otimes X$ acts on $\lceil\log k\rceil+1$ qubits, and $M_y = \ketbra{y} + (1-\eta)(\mathbf{I}-\ketbra{y})$, we arrive at the following decomposition:
\begin{eqnarray}
    U_y&=& \ketbra{y}\otimes Z+(I-\ketbra{y})\otimes((1-\eta)Z+\sqrt{2\eta-\eta^2}X)\\
    &=& [I\otimes((1-\eta)Z+\sqrt{2\eta-\eta^2}X)][\ketbra{y}\otimes ((1-\eta)Z+\sqrt{2\eta-\eta^2}X)Z+(I-\ketbra{y})\otimes I]. \nonumber
\end{eqnarray}
The first part of the above equation $I\otimes((1-\eta)Z+\sqrt{2\eta-\eta^2}X)$ is a single-qubit gate and can be compiled into a constant number of gates. The second part is a $\lceil\log k\rceil$-controlled $SU(2)$ gate, as $\ketbra{y}$ involves $\lceil\log k\rceil$ qubits and $(1-\eta)Z+\sqrt{2\eta-\eta^2}X)Z$ is a $SU(2)$ gate. Utilizing the techniques in Ref. \cite{valeDecompositionMulticontrolledSpecial2023}, such a multi-controlled $SU(2)$ gate can be compiled into $O(\log k)$ CNOT gates. This leads to the conclusion that the whole $U_y$ can be compiled in an efficient way with about  $\log k$ two-qubit gates.

\subsection{Prediction and evaluataion}

The predicted label of $\bx$ is the outcome of measurements in the computational basis performed on $U(\bx)\ket{\psi}$. So the probability of correct prediction (i.e., the accuracy) is $\braket{\psi|U(\bx)^{\dagger}\Pi_{f(\bx)}U(\bx)|\psi}=1-\braket{\psi|H_{\bx}|\psi}$. This accuracy can be amplified by repetition. Indeed, once the data in each step has been sampled, the training process is a fixed quantum circuit (with post-selection). So we can run the circuit multiple times to obtain $K$ copies of the final states $\ket{\psi}$. To predict the label of a new unseen data  sample $\bx$, we measure  $U(\bx)\ket{\psi}$ in the computational basis for each copy and do a majority vote. For simplicity, we consider the binary classification problem and assume $K$ is odd, then the probability of correct label (called $K$-accuracy) is 
\begin{equation*}
    p_K(\bx, \ket{\psi}) = \sum_{r=0}^{(K-1)/2}\binom{K}{r} \braket{\psi|H_{\bx}|\psi}^r (1-\braket{\psi|H_{\bx}|\psi})^{K-r}.
\end{equation*}
When $K=1$, this reduces to the single-copy accuracy $1-\braket{\psi|H_{\bx}|\psi}$. When $K= \infty$, the $K$-accuracy equals to the step function $1[\braket{\psi|H_{\bx}|\psi}<1/2]$. Throughout this paper, we call $K$ the number of trials.

\subsection{Gradient perspective}

As mentioned in the main text, the dissipation process of quantum automated learning actually implements the gradient descent algorithm in an automated manner. 
In the training step (iii), the quantum system is evolved through $U(\bx)$, $M_{y}$ and $U(\bx)^\dagger$. We define $M_{y} = \Pi_{y(\bx)} + (1-\eta) (\mathbf{I}-\Pi_{y(\bx)})$ and $H_{\bx}=\mathbf{I}-U(\bx)^\dagger \Pi_{y(\bx)}U(\bx)$, where $\Pi_{y(\bx)}$ denotes the measurement projection corresponding to the state encoding the label $y$. Then we get the updated (unnormalized) state $U(\bx)^\dagger M_{y} U(\bx) |\psi\rangle = (\mathbf{I} - \eta H_{\bx}) |\psi\rangle$. 

As mentioned above, the non-unitary perturbation $M_{y}$ in training step (iii) of the QAL protocol is implemented by block encoding into a unitary with an ancillary qubit combined with post-selection. As a result, this step effectively updates the state with unitary transformation: 
\begin{equation}\label{equ:update}
    \ket{\psi}\leftarrow \frac{(I-\eta H_{\bx})\ket{\psi}}{\norm{(I-\eta H_{\bx})\ket{\psi}}},
\end{equation} 
where $\norm{(I-\eta H_{\bx})\ket{\psi}}$ is a normalization factor whose square gives the success probability of post-selection.

The probability of correct prediction of a datum $\bx$ reads $\braket{\psi|U(\bx)^{\dagger}\Pi_{y(\bx)}U(\bx)|\psi}=1-\braket{\psi|H_{\bx}|\psi}$. 
As mentioned in the main text, we define the loss function as the average failure probability: $\hat{R}_S(\psi) = \E_{\bx\sim S}\braket{\psi|H_{\bx}|\psi}$,

where $\E_{\bx\sim S}$ denotes the expectation and 
$\bx\sim S$ means $\bx$ is uniformly sampled from the training set $S$. 
From the perspective of conventional machine learning, we may also regard $\ket{\psi}$ as a variational state parametrized by a complex vector $\bm{\psi}$. 
Given that the expectation value $\braket{\psi|H_{\bx}|\psi}$ is real, we transform the complex vector $\bm{\psi} = (\psi_1,\dots,\psi_{2^n})$ into a fully real representation: $(a_1,b_1,\dots,a_{2^n},b_{^n})$, where $a_i$ and $b_i$ denote the real and imaginary components of $\psi_i$ respectively. Owing to the Hermitian property of $H_x$, we derive the partial derivatives: $\frac{\partial \braket{\psi|H_{\bx}|\psi}}{\partial a_i} = 2Re(\sum_j(H_x)_{i,j}\psi_j)$ and $\frac{\partial \braket{\psi|H_{\bx}|\psi}}{\partial b_i} = 2Im(\sum_j(H_x)_{i,j}\psi_j)$. This allows us to define $\frac{\partial \braket{\psi|H_{\bx}|\psi}}{\partial \psi_i} = 2\sum_j(H_x)_{i,j}\psi_j$. Consequently, the gradient of $\braket{\psi|H_{\bx}|\psi}$ with respect to $\psi$ can be succinctly expressed as $2H_{\bx}\ket{\psi}$.

Therefore, the update rule  in Eq. \eqref{equ:update} essentially implements the stochastic projected gradient descent algorithm to minimize the loss function $\hat{R}_S(\psi)$ with a batch size one. Here we use the term ``projected'' to emphasize the normalization after each update. 
From the stochastic gradient descent perspective, one may conclude that an initial state $|\psi\rangle$ can exponentially converge to a local minimum through updating rule \eqref{equ:update} on expectation \cite{mohri2018foundations}.  However, a rigorous proof of convergence to the global minimum is unattainable in general. In fact, this is an inherent drawback for conventional gradient-based quantum learning approaches. Whereas, owing to the quadratic form of the loss function and the clear physical interpretation, we can rigorously prove that  $\ket{\psi}$ converges exponentially to the global minimum for the QAL protocol, as discussed in the main text and detailed in the following sections.

\section{Physical Interpretation and Analytical Results}
Throughout this section, we use $\norm{A}_1, \norm{A}_\infty$ to denote the trace norm (the summation of singular values) and spectral norm (the largest singular value), respectively. For two Hermitian $A, B$, denote $A\preceq B$ if $B-A$ is positive semi-definite. By definition, for any $\bx$, $0\preceq H_{\bx}\preceq I$, and thus $0\preceq H_{S}\preceq I$. We will use the following fact.
\begin{lemma}\label{lem:norm_inequality}
    Let $A, B, C$ be three Hermitian matrices such that $\norm{A}_1\leq 1$, $0\preceq B, C\preceq I$. Then
    \begin{equation}
        \norm{BAB}_1\leq 1, \norm{BAC+CAB}_1\leq 2. 
    \end{equation}
\end{lemma}
\subsection{Formulation of the training process}
In this subsection, we explain the training process from a physical perspective and derive an analytical characterization of the success probability of post-selection and the performance of the final model. Observe that the empirical risk 
is the energy of $\ket{\psi}$ under the Hamiltonian $H_S$, so finding the global minimum is equivalent to finding the ground state of $H_S$. We rewrite \eqref{equ:update} in the density matrix formalism:
\begin{equation}\label{equ:update_density}
    \rho\xleftarrow{\bx} (I-\eta H_{\bx})\rho (I-\eta H_{\bx}).
\end{equation}
Here we keep the post-state $\rho$ unnormalized. Indeed, $\Tr(\rho)$ is the success probability of the post-selection. So the density matrix formalism helps us to keep track of the overall success probability. Another benefit of the density matrix formalism is that we can embed the randomness of the sample into the state. Since the datum $\bx$ is uniformly sampled from $S$, the averaged post-state up to the second order term is 
\begin{align}
    \rho &\leftarrow \E_{\bx\sim S} (I-\eta H_{\bx})\rho (I-\eta H_{\bx}) \nonumber\\
    &\approx I-\eta (H_S \rho+\rho H_S)\nonumber\\
    &\approx e^{-\eta H_S}\rho e^{-\eta H_S}. \label{equ:imaginary_time}
\end{align}
We make this approximation precise in the following lemma
\begin{lemma}\label{lem:imaginary_time_single_step}
    \begin{equation}
        \E_{\bx\sim S} (I-\eta H_{\bx})\rho (I-\eta H_{\bx}) = e^{-\eta H_S}\rho e^{-\eta H_S} + \eta^2 O,
    \end{equation}
    where $O$ is a Hermitian matrix with trace norm at most 4. 
\end{lemma}
\begin{proof}
    Let $R=(e^{-\eta H_S}-(I-\eta H_S))/\eta^2$. Since $0\preceq H_S\preceq I$, all the eigenvalues of $H_S$ are in $[0, 1]$. By the Taylor expansion of the exponential function, for any $x\in [0, 1]$, there exists $x^*\in [0, 1]$ such that $(e^{-\eta x}-(1-\eta x))/\eta^2 = (x^*)^2/2\in [0, 1/2]$. Therefore, $R$ is a Hermitian matrix such that $0\preceq R\preceq I/2$. By \Cref{lem:norm_inequality}, we have
    \begin{align}
        &\norm{\E_{\bx\sim S} (I-\eta H_{\bx})\rho (I-\eta H_{\bx}) - e^{-\eta H_S}\rho e^{-\eta H_S}}_1\nonumber\\
        =& \norm{\rho-\eta (H_S\rho +\rho H_S) + \eta^2 \E_{\bx\sim S}H_{\bx} \rho H_{\bx} - (\eta^2 R + (I-\eta H_S))\rho (\eta^2 R + (I-\eta H_S))}_1\nonumber\\
        =& \eta^2\norm{\big(\E_{\bx\sim S}H_{\bx}\rho H_{\bx} -H_S\rho H_S -\eta^2 R\rho R -(R\rho (I-\eta H_S)+(I-\eta H_S)\rho R)\big)}_1\nonumber\\
        \leq &\eta^2(1+1+\eta^2/4 +1) < 4\eta^2.\qedhere
    \end{align}
\end{proof}
Up to the second order term, \eqref{equ:imaginary_time} is the imaginary time evolution of $\rho$ under $H_S$. Suppose the initial state is $\sigma$, then the averaged state after $T$ epochs is $\rho = e^{-\eta T H_S}\sigma e^{-\eta T H_S}$. Let $\beta=\eta T$ be the summation of learning rates. We can approximate the success probability of possibility by $\tr(e^{-\beta H_S}\sigma e^{-\beta H_S})$ and the loss by $\tr(H_S\frac{e^{-\beta H_S}\sigma e^{-\beta H_S}}{\tr(e^{-\beta H_S}\sigma e^{-\beta H_S})})$. We summarize and prove the results in the following theorem.

\begin{theorem}\label{thm:formulation}
    Suppose we train the QAL model with initial state $\sigma$ for $T$ steps, with learning rate $\eta_t$ at step $t$. Define
    \begin{equation}
        \beta = \sum_{i=1}^T \eta_t,\quad \gamma = \sum_{i=1}^T \eta_t^2, \quad \sigma(\beta)=e^{-\beta H_S}\sigma e^{-\beta H_S}.
    \end{equation}
    Averaging over choice of training samples, the success probability of post-selection is 
    \begin{equation}
        \tr(\sigma(\beta))+c_1\gamma,\label{eq:success_probability}
    \end{equation}
    and the average loss conditioned on the success of post-selection is
    \begin{equation}
        \frac{\tr(H_S\sigma(\beta))+c_2\gamma}{\tr(\sigma(\beta))+c_1\gamma}. \label{eq:conditional_loss}
    \end{equation}
    Here $c_1, c_2$ are two real numbers such that $\abs{c_1}, \abs{c_2}\leq 4$.
\end{theorem}
\begin{proof}
    Let $\bx_1, \cdots, \bx_T \sim S$ be the training samples in the $T$ steps. We abbreviate $\bx_1, \cdots, \bx_t$ as $\bx_{1:t}$.
    By~\eqref{equ:update_density}, the unnormalized state after step $t$ is
    \begin{equation}
        \rho_t^{\bx_{1:t}} = (I-\eta_t H_{\bx_t})\cdots (I-\eta_1 H_{\bx_1})\sigma (I-\eta_1 H_{\bx_1})\cdots (I-\eta_t H_{\bx_t}).
    \end{equation}
    Given samples $\bx_{1:T}$, $\tr(\rho_T^{\bx_{1:T}})$ is the success probability of post-selection and $\tr(H_S\frac{\rho_T^{\bx_{1:T}}}{\tr(\rho_T^{\bx_{1:T}})})$ is the loss conditioned on the success of post-selection. We now average over the choice of samples.
    Recursively apply Lemma \ref{lem:imaginary_time_single_step} and Lemma \ref{lem:norm_inequality}, we have 
    \begin{equation}
        \E_{\bx_{1:T}\sim S^T}\rho_T^{\bx_{1:T}} = \sigma(\beta) + \gamma O,
    \end{equation}
    for some Hermitian $O$ with trace norm at most 4. The average success probability of post-selection is 
    \begin{equation}
        \E_{\bx_{1:T}\sim S^T}\tr(\rho_T^{\bx_{1:T}}) = \tr(\sigma(\beta)) + c_1\gamma,
    \end{equation}
    where $c_1=\tr(O)$ satisfies $\abs{c_1}\leq 4$. Now we calculate the averaged loss conditioned on the success of post-selection. For clarity, denote $q=\tr(\sigma(\beta)) + c_1\gamma$, $p=1/\abs{S}^T$ be the probability of sampling $\bx_1, \cdots, \bx_T$.
    Then conditioned on the success of post-selection, the conditional probability of sampling $\bx_{1:T}$ is $p\tr(\rho_T^{\bx_{1:T}})/q$. Therefore, the average loss conditioned on the success of post-selection is
    \begin{equation}
        \sum_{\bx_{1:T}\sim S^T} \frac{p\tr(\rho_T^{\bx_{1:T}})}{q} \tr(H_S\frac{\rho_T^{\bx_{1:T}}}{\tr(\rho_T^{\bx_{1:T}})}) = \frac{1}{q}\E_{\bx_{1:T}\sim S^T}\tr(H_S\rho_T^{\bx_{1:T}})=\frac{\tr(H_S\sigma(\beta)) + c_2\gamma}{\tr(\sigma(\beta)) + c_1\gamma},
    \end{equation}
    where $c_2=\tr(H_S O)$ satisfies $\abs{c_2}\leq 4$.
\end{proof}

According to the theorem, up to the second order term $c_1\gamma, c_2\gamma$, the training process behaves the same as the imaginary time evolution of $\sigma$ under $H_S$. The effect of imaginary time evolution is clearer in the eigenbasis of $H_S$. Write the spectrum decomposition of $H_S$ as $H_S=\sum_{i}E_i\ketbra{E_i}$ and define $\sigma_i=\braket{E_i|\sigma|E_i}$ as the overlap of $\sigma$ with the $i$-th eigenstate. Then 
\begin{equation}
    \sigma(\beta)=\sum_{i}\sigma_i e^{-2\beta E_i}\ketbra{E_i},~\quad
    \frac{\tr(H_S\sigma(\beta))}{\tr(\sigma(\beta))} = \frac{\sum_i E_i\sigma_i e^{-2\beta E_i}}{\sum_i \sigma_i e^{-2\beta E_i}}.
\end{equation}
The weight of $\ketbra{E_i}$, $\sigma_i e^{-2\beta E_i}$, decays exponentially with $\beta$. The decay is slower for lower energy eigenstates. Assume $\sigma$ has a non-zero overlap with the ground space. As $\beta$ goes up, eventually the weight of the ground space dominates, so $\rho_{\text{n}}(\beta)$ converges to a ground state of $H_S$ and the empirical risk converges to the global minimum. In the following, we will make this intuition rigorous in the presence of $c_1\gamma, c_2\gamma$. 

\subsection{Convergence to global minimum}
Denote the ground energy of $H_S$ (i.e., the global minimum of the loss) as $g$, the projector to the ground space as $\Pi_g$, and the gap between the ground energy and the first excited state as $\delta>0$. 
\begin{theorem}\label{thm:convergence_to_global_minimum}
    Suppose $\sigma$ has a nonzero overlap with the ground space of $H_S$ (that is, $\sigma_g=\tr(\Pi_g\sigma)>0$). For any constant $c\in (0, 1)$, we can choose an appropriate $\eta$ and $T$ such that if we train the QAL model for $T$ steps with learning rate $\eta$ in each step, the averaged loss conditioned on the success of post-selection is at most $g+c$.
    
\end{theorem}
\begin{proof}
    According to Theorem~\ref{thm:formulation}, we only need to upper bound~\eqref{eq:conditional_loss} for $\beta = \eta T$ and $\gamma=\eta^2 T=\beta\eta$. Since
    \begin{align}
        \frac{\tr(H_S\sigma(\beta))+c_2\beta\eta}{\tr(\sigma(\beta))+c_1\beta\eta}&\leq \frac{\sigma_g e^{-2\beta g}g + (1-\sigma_g)e^{-2\beta (g+\delta)}+4\beta\eta}{\sigma_g e^{-2\beta g} -4\beta\eta} \nonumber\\
        &=g + \frac{(1-\sigma_g)e^{-2\beta (g+\delta)}+(4+4g)\beta\eta}{\sigma_g e^{-2\beta g} -4\beta\eta}\nonumber\\
        &\leq g + \frac{e^{-2\beta \delta}+8\beta\eta e^{2\beta g}}{\sigma_g - 4\beta\eta e^{2\beta g}}\label{eq:global_minimum_1}.
    \end{align}
    Choose $\beta$ such that $e^{-2\beta\delta}<\sigma_gc/4$, and then choose $\eta$ such that $\beta\eta e^{2\beta g}<\sigma_gc/16<\sigma_g/16$. Then the right hand side of~\eqref{eq:global_minimum_1} is at most
    \begin{equation}
        g+\frac{\sigma_g c/4+\sigma_g c/2}{\sigma_g-\sigma_g/4} = g+c.\qedhere
    \end{equation}
\end{proof}

A randomly initialized state $\sigma$ has a nonzero overlap with the ground space with probability 1. According to the theorem, the QAL model will converge to the global minimum of the loss function. However, this convergence is built on the success of post-selection, whose probability exponentially decays with the number of steps. Therefore, a more realistic question is whether we can build a reasonable trade-off between the success probability and the performance of the final model. 

\subsection{Convergence with constant probability}
In this subsection, we will establish a practical trade-off between the accuracy of the final model and the success probability of post-selection when the initial state has a large overlap with the low-energy eigenspace of $H_S$.

\begin{definition}
    Let $H$ be a Hamiltonian. The $E$ low energy subspace of $H$ is the subspace spanned by the eigenstates of $H$ with energy at most $E$. Denote the projector to the $E$ low energy subspace as $\Pi_E^H$. The overlap of a state $\sigma$ with the $E$ low energy subspace is defined as $\tr(\Pi_E^H\sigma)$.
\end{definition}

Throughout this section, we focus on the Hamiltonian $H_S$ and omit the superscript $H$. 

\begin{theorem}\label{thm:constant_probability}
    Let $c_1, c_3\in (0, 1), c_2\in (0, 1/10)$ be three constants, $g$ be the ground energy of $H_S$, and $\epsilon>0$ such that $g/\epsilon \leq c_1$. Assume the overlap between the initial state $\sigma$ and the $(g+\epsilon)$ low energy eigenspace of $H_S$, namely $\tr(\sigma \Pi_{g+\epsilon})$, is at least $c_2$. Then we can choose an appropriate $\eta$ and $T$ such that if we train the QAL model with the initial state $\sigma$ for $T$ steps with learning rate $\eta$ in each step, the success probability of post-selection is at least $c_4$ and the averaged loss conditioned on the success of post-selection is at most $g+\epsilon+c_3$. Here $c_4$ is a constant that only depends on $c_1, c_2, c_3$. 
\end{theorem}
\begin{proof}
    By Theorem~\ref{thm:formulation}, we only need to lower bound the the success probability in \eqref{eq:success_probability} and the conditional loss in \eqref{eq:conditional_loss}. We will follow the notation in Theorem~\ref{thm:formulation}, so that $\beta = \eta T$, $\gamma = \eta^2 T$, and $\sigma(\beta)=e^{-\beta H}\sigma e^{-\beta H}$. Here we write $H=H_S$ for simplicity. We will prove the theorem for $\beta = 3\ln(1+c_2)/(c_3\epsilon)$, $c_4=e^{-6(1+c_1)\ln(1+c_2)/c_3}c_2/2$ and $\gamma=c_3c_4/40$. Accordingly, $\eta = \gamma/\beta$ is of order $\epsilon$ and $T=\beta^2/\gamma$ is of order $1/\epsilon^2$. 
    By~\eqref{eq:success_probability}, the success probability is at least
    \begin{align}
        \tr(\sigma(\beta)) - 4\gamma &\ge \tr(\Pi_{g+\epsilon}\sigma(\beta))-4\gamma\nonumber \\
        &=\tr(e^{-\beta H}\Pi_{g+\epsilon}e^{-\beta H}\sigma)-4\gamma\nonumber\\
        &\ge e^{-2\beta(g+\epsilon)}\tr(\Pi_{g+\epsilon}\sigma)-4\gamma\nonumber\\
        &\ge e^{-2\beta\epsilon(1+c_1)}c_2-4\gamma\nonumber\\
        &= 2c_4-4\gamma > c_4.
    \end{align}
    By~\eqref{eq:conditional_loss}, the averaged loss conditioned on the success of post-selection is at most
    \begin{align}
        \frac{\tr(H \sigma(\beta))+4\gamma}{\tr(\sigma(\beta))-4\gamma}&=g+\frac{\tr((H-gI)\sigma(\beta))}{\tr(\sigma(\beta))-4\gamma}+\frac{(4+4g)\gamma}{\tr(\sigma(\beta))-4\gamma}\nonumber\\
        &\leq g + \frac{\tr((H-gI)\sigma(\beta))}{\tr(\sigma(\beta))(1-c_3/20)}+\frac{8\gamma}{c_4}\nonumber\\
        &\leq g + \frac{c_3}{5} + (1+\frac{c_3}{5})\frac{\tr((H-gI)\sigma(\beta))}{\tr(\sigma(\beta))},\label{eq:constant_probability_proof1}
    \end{align}
    where we use $4\gamma=c_3c_4/10\leq c_3\tr(\sigma(\beta))/20$ in the second line and $(1+c_3/5)(1-c_3/20)\ge 0$ in the third line.
    So it suffices to upper bound $\tr((H-gI)\sigma(\beta))/\tr(\sigma(\beta))$. 
    Write the spectrum decomposition of $H$ as $H=\sum_i E_i\ketbra{E_i}$ and let $x_i = 2\beta(E_i-g), \sigma_i=\braket{E_i|\sigma|E_i}$. We simplify the last term of~\eqref{eq:constant_probability_proof1} to
    \begin{align}
        \frac{\tr((H-gI)\sigma(\beta))}{\tr(\sigma(\beta))} &= \frac{\sum_i (E_i-g)e^{-2\beta E_i}\sigma_i}{\sum_i e^{-2\beta E_i}\sigma_i}\nonumber\\
        &=\frac{\sum_{i}\sigma_i e^{-x_i}x_i}{\sum_{i}\sigma_i e^{-x_i}}\cdot \frac1{2\beta}.\label{eq:constant_probability_proof2}
    \end{align}
    Split the Hilbert space into low energy and high energy eigenspaces, $L=\{i: E_i\leq g+\epsilon\}$ and $H=\{i: E_i>g+\epsilon\}$. Let $p_L=\tr(\sigma \Pi_{g+\epsilon})=\sum_{i\in L}\sigma_i\ge c_2$ and $p_H=1-p_L$ be the overlaps of $\sigma$ with the two subspaces. 
    Since $f(y)=-y\ln(y)$ is concave, by Jensen's inequality, 
    \begin{equation}
        \sum_{i\in L}\frac{\sigma_i}{p_L} f(e^{-x_i})\leq f\big(\sum_{i\in L}\frac{\sigma_i}{p_L}e^{-x_i}\big).
    \end{equation}
    Let $l$ be the number such that $e^{-l}=\sum_{i\in L}\sigma_i e^{-x_i}/p_L$. The inequality becomes $\sum_{i\in L}\sigma_i e^{-x_i}x_i\leq p_L e^{-l}l$. Similarly, let $h=-\ln(\sum_{i\in H}\sigma_i e^{-x_i}/\sum_{i\in H}\sigma_i)$, then $\sum_{i\in H}\sigma_i e^{-x_i}x_i\leq p_He^{-h}h$. Therefore, 
    \begin{equation}
        \frac{\sum_{i}\sigma_i e^{-x_i}x_i}{\sum_{i}\sigma_i e^{-x_i}} \leq \frac{p_L e^{-l}l+p_He^{-h}h}{p_L e^{-l}+p_He^{-h}}.\label{eq:constant_probability_proof3}
    \end{equation}
    By definition, $x_i\in [0, 2\beta \epsilon]$ for $i\in L$ and $x_i>2\beta \epsilon $ for $i \in H$. Since $e^{-l}$ is a mixed of $e^{-x_i} (i\in L)$, we have $0\leq l\leq 2\beta\epsilon$ and similarly $h\ge 2\beta\epsilon$. Denote $y=h-l$. Insert~\eqref{eq:constant_probability_proof2} and~\eqref{eq:constant_probability_proof3} to~\eqref{eq:constant_probability_proof1}.
    \begin{align}
        \frac{\tr(H \sigma(\beta))+4\gamma}{\tr(\sigma(\beta))-4\gamma}&\leq g+\frac{c_3}{5}+ (1+\frac{c_3}{5})\frac{p_L e^{-l}l+p_He^{-h}h}{p_L e^{-l}+p_He^{-h}}\cdot\frac{1}{2\beta}\nonumber\\
        &= g + \frac{c_3}{5}+(1+\frac{c_3}{5})(l+\frac{p_H e^{-y}y}{p_L+p_He^{-y}})\cdot \frac1{2\beta}\nonumber\\
        &\leq g+\frac{c_3}{5} + (1+\frac{c_3}{5})(l+\frac{e^{-y}y}{c_2+e^{-y}})\cdot \frac1{2\beta}. \label{eq:constant_probability_proof4}
    \end{align}
    By differentiating $g(y)=e^{-y}y/(c_2+e^{-y})=y/(c_2e^{y}+1)$, we find that $g(y)\leq g(y^*)$ for the $y^*>0$ such that $c_2e^{y^*}(y^*-1)=1$. For this $y^*$ we have $g(y^*)=y^*-1$. Assume $y^*>\ln(1/c_2)>2$, then $1=c_2e^{y^*}(y^*-1)> c_2(1/c_2)(2-1)=1$, a contradiction. So $g(y)\leq g(y^*)=y^*-1\leq \ln(1/c_2)$. By~\eqref{eq:constant_probability_proof4}, the averaged loss conditioned on the success of post-selection is at most
    \begin{equation}
        g+\frac{c_3}{5} + (1+\frac{c_3}{5})(2\beta\epsilon+\ln(1/c_2))\cdot \frac1{2\beta} = g+\frac{c_3}{5}+(1+\frac{c_3}{5})(\epsilon+\frac{c_3\epsilon}{6})<g+\epsilon+c_3.\qedhere
    \end{equation}
\end{proof}

The theorem ensures the convergence of the QAL training process with a constant success probability, assuming a good initial state. Concretely, as long as the initial state has a large (at least $c_2$) overlap with the low energy eigenspace (with energy at most $g+\epsilon$), the QAL training process will converge to the low energy eigenspace (up to an arbitrarily small residue error $c_3$) with a constant success probability ($c_4$ that only depends on $c_1, c_2, c_3$). 

\subsection{Heavy-tailed Hamiltonian}
Theorem \ref{thm:constant_probability} highlights the importance of the initial state $\sigma$. However, without prior knowledge of $H_S$, we cannot do better than a random guess, or equivalently, starting from the maximally-mixed state $\sigma = I/2^n$. Therefore, we actually hope that $H_S$ has a constant proportion of low energy eigenstates that do not scale up with $n$, the dimension of $\bx$, and the size $m$ of the training dataset. We formalize this intuition in the following definition.

\begin{definition}[Heavy-tailed Hamiltonian]
    We say a Hamiltonian $H$ is $(E, c)$-heavy-tailed if the proportion of eigenstates with energy at most $E$ is at least $c$.
\end{definition}

\begin{theorem}\label{thm:heavy_tailed}
    Let $c_1, c_3\in (0, 1), c_2\in (0, 1/10)$ be three constants. Suppose $H_S$ is $(g+\epsilon, c_2)$-heavy-tailed, where $g$ is the ground energy of $H_S$ and $\epsilon>0$ such that $g/\epsilon \leq c_1$. Then we can choose an appropriate $\eta$ and $T$ such that if we train the QAL model with a maximally-mixed initial state in computational basis for $T$ steps with learning rate $\eta$ in each step, the success probability of post-selection is at least $c_4$ and the averaged loss conditioned on the success of post-selection is at most $g+\epsilon+c_3$. Here $c_4$ is a constant that only depends on $c_1, c_2, c_3$.
\end{theorem}

\begin{proof}
    By definition of heavy-tailed Hamiltonian, the overlap between the initial state $\sigma=I/2^n$ and the $(g+\epsilon)$ low eigenspace of $H_S$ is at least $c_2$. The theorem follows directly from Theorem~\ref{thm:constant_probability}.
\end{proof}

Therefore, the QAL training process is guaranteed to converge to the low energy eigenspace of a heavy-tailed $H_S$.
We now argue that when $H_S$ comes from a reasonable dataset, it is likely to be heavy-tailed due to the similarity of data.
Consider the extremely simple example of classifying dogs and cats, where all dogs look similar and all cats look similar. The Hamiltonian $H_S$ is approximately a mixture of two projectors of dimensions $2^{n-1}$, $H_{\text{dogs}}$ and $H_{\text{cats}}$. Regard $H_{\text{dogs}}$ and $H_{\text{cats}}$ as random projectors, then $H_S$ has a constant proportion of near-zero eigenvalues. This assumption is supported by the numerical simulation.

Remark that while Theorem~\ref{thm:heavy_tailed} applies to the maximally-mixed initial state, in reality we will use a random initial state in the computational basis. Once we sample an initial state better than the maximally-mixed state, we can stick to it and apply Theorem~\ref{thm:constant_probability}.

\subsection{Generalization}
The previous results establish the explainable trainability of QAL. While in training classical neural networks, each epoch is a full pass of the training dataset, in QAL, each step only involves a single datum. This indicates that the QAL model could be optimized using a few data points. In this subsection we rigorously demonstrate the generalization ability of QAL, showing that once the model achieves a good performance on a small dataset, the good performance will generalize to unseen data.

Recall that the training loss and the true loss of $\ket{\psi}$ are $\hat{R}_S(\psi)=\E_{x\sim S}\braket{\psi|H_{\bx}|\psi}$ and $R(\psi)=\E_{\bx\sim \cD}\braket{\psi|H_{\bx}|\psi}$, respectively.
\begin{theorem}\label{thm:generalization}
    With probability at least $1-\delta$ over the choice of $S$, the generalization gap is upper bounded by
    \begin{equation}
        \max_{\ket{\psi}}\big(R(\psi)-\hat{R}_S(\psi)\big) \leq \sqrt{\frac{4\ln(2^{n+1}/\delta)}{m}}.
    \end{equation}
\end{theorem}
\begin{proof}
    By definition, the left-hand side is upper bounded by the spectral norm of $\E_{x\sim_u S}H_{x}-\E_{x\sim \cD}H_{x}$, which can be bounded by matrix Bernstein inequality (see, e.g., \cite[Theorem 6.1.1]{troppIntroductionMatrixConcentration2015}): \begin{equation*}
        \Pr_S[\norm{\E_{x\sim_u S}H_x-\E_{x\sim \cD}H_x}_2\ge t]\leq 2^{n+1}\exp(-mt^2/4).
    \end{equation*}
    The theorem follows by setting $t=\sqrt{4\ln(2^{n+1}/\delta)/m}$.
\end{proof}
According to the theorem, as long as the size $m$ of the training dataset is larger than $\Omega(n)$ (i.e., a logarithm of the degree of freedom), a parameter state $\ket{\btheta}$ with low training loss has a low true loss with high probability. This is better than quantum neural networks where the training dataset size has to be larger than the degree of freedom \cite{caiSampleComplexityLearning2022,caroGeneralizationQuantumMachine2022}. 
We remark that the good generalization stems from the simple quadratic form of the loss function.

\section{More Numerical Results}
\begin{figure}
    \centering
    \includegraphics[width=\linewidth]{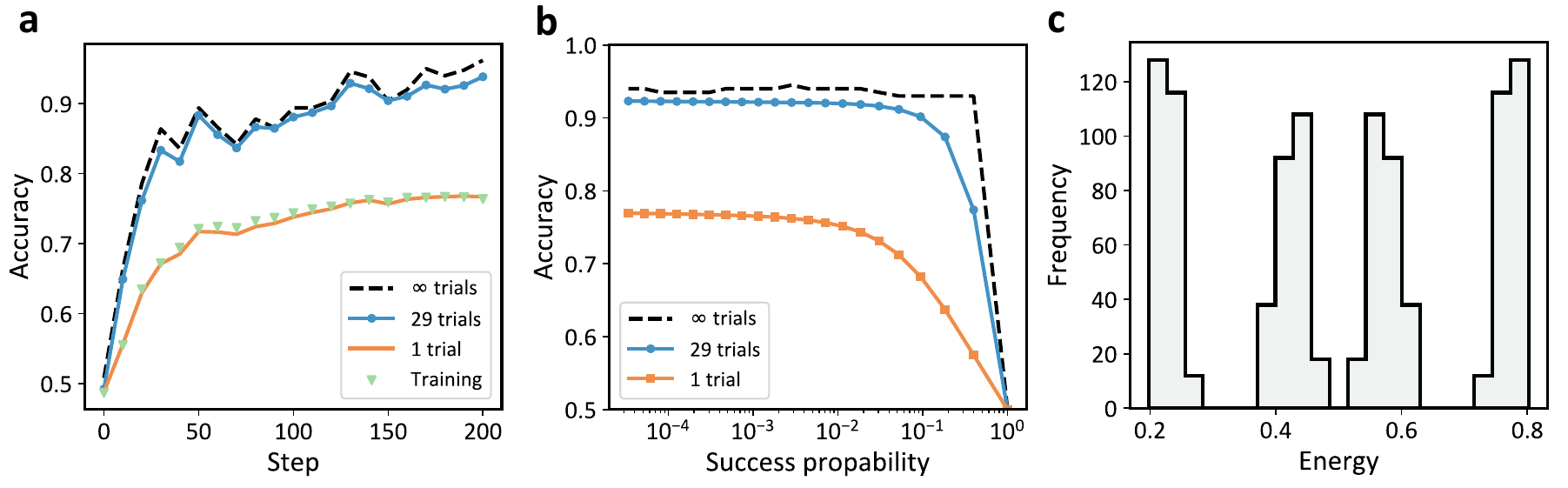}
    \captionsetup{justification=raggedright, singlelinecheck=false}
    \caption{Classify images in the MNIST dataset \cite{lecun1998mnist} using QAL. \textbf{a}, Testing and training accuracy during the training process. \textbf{b} The trade-off between testing accuracy and the success probability of post-selection. \textbf{c}, Spectrum of the associated Hamiltonian $H_S$ of the training dataset.}
    \label{fig:mnist}
\end{figure}

\begin{figure}
    \centering
    \includegraphics[width=\linewidth]{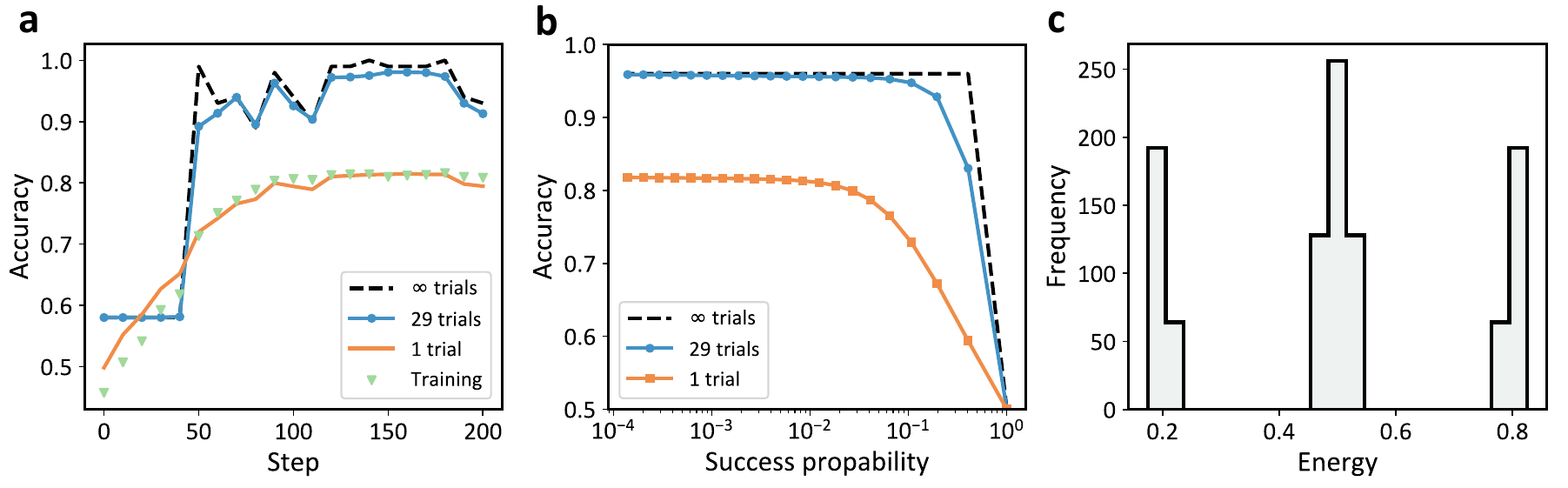}
    \captionsetup{justification=raggedright, singlelinecheck=false}
    \caption{Classify Aubry-Andr\'{e} Hamiltonian \cite{aubry1980analyticity} using QAL. \textbf{a}, Testing and training accuracy during the training process. \textbf{b} The trade-off between testing accuracy and the success probability of post-selection. \textbf{c}, Spectrum of the associated Hamiltonian $H_S$ of the training dataset.}
    \label{fig:AA}
\end{figure}

\begin{figure}
    \centering
    \includegraphics[width=\linewidth]{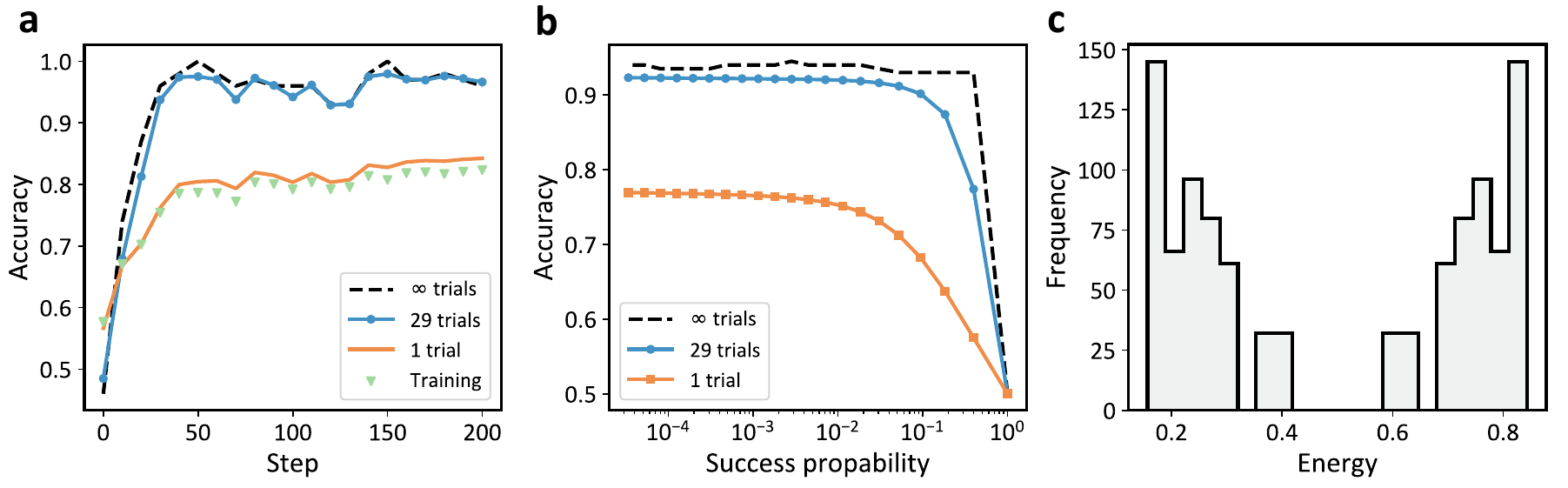}
    \captionsetup{justification=raggedright, singlelinecheck=false}
    \caption{Classify the ground state of clustering-Ising model  \cite{son2011quantum,smacchia2011statistical} using QAL. \textbf{a}, Testing and training accuracy during the training process. \textbf{b} The trade-off between testing accuracy and the success probability of post-selection. \textbf{c}, Spectrum of the associated Hamiltonian $H_S$ of the training dataset.}
    \label{fig:CIM}
\end{figure}

Besides the numerical simulation on the Fashion MNIST dataset in the main text, we also conduct QAL protocol on MNIST dataset, Hamiltonian dataset, and quantum state dataset. The data encoding scheme for each dataset is describe in For each dataset. We keep the setting the same, so that the number of qubits is 10, the learning rate is $0.1$, and the test data size is 500. For each dataset, we plot the training performance, trade-off between accuracy and success probability of post-selection, and spectrum of data Hamiltonian as in the main text.

MNIST is a widely used benchmark dataset for handwritten digit recognition~\cite{lecun1998mnist}. Here, we follow the same data processing routine as Fashion MNIST. We focus on the binary classification of classes "1" and "9". All images are rescaled to $10\times 10$ pixels, and pixel values are normalized. The results are shown in Fig.~\ref{fig:mnist}.

For Hamiltonian data, we consider a binary classification of the following Aubry-Andr\'{e} Hamiltonian on 10 qubits~\cite{aubry1980analyticity}:
\begin{equation*}
H=-\frac g2 \sum_k (\sigma_k^x\sigma_{k+1}^x+\sigma_k^y\sigma_{k+1}^y)-\frac V2 \sum_k \cos(2\pi \phi k)\sigma_k^z, 
\end{equation*}
where $g$ is the coupling strength, $\sigma_k^\alpha$ ($\alpha=x, y, z$) are Pauli operators on the $k$-th qubit, $V$ is the disorder magnitude and $\phi=(\sqrt{5}-1)/2$. This Hamiltonian exhibits a quantum phase transition at $V/g=2$, between a localized phase for $V/g>2$ and a delocalized phase for $V/g<2$. To generate the dataset, we fix $g=1$, sample $V$ in the interval $[0, 4]$ and label the Hamiltonian according to its phase. To carry out the QAL protocol, we encode the Hamiltonian into its real-time evolution $e^{-2iH}$. Our numerical results for Hamiltonian data are ploted in Fig.~\ref{fig:AA}.

For quantum state data, we classify the ground state of 10-qubit clustering-Ising model Hamiltonian~\cite{son2011quantum,smacchia2011statistical} with periodic boundary condition:
\begin{equation*}
    H(h)=-\sum_{k}\sigma_k^x\sigma_{k+1}^z\sigma_{k+2}^x+h\sum_{k}\sigma_k^y\sigma_{k+1}^y,
\end{equation*}
This Hamiltonian has a phase transition at $h=1$, between a symmetry protected topological phase ($h<1$) and an antiferromagnetic phase ($h>1$). The dataset is generated by sampling $h$ from interval $[0, 2]$ and labeling the state according to its phase. As mentioned in Sec.~\ref{supp_sec:data_encoding}, the data encoding scheme relies on a Hamiltonian. In our numerical simulations, $H$ is a random 4-local Hamiltonian. Each term of $H$ is a 4-body Pauli interaction on 4 random positions with a random interaction strength between -1 and 1. The results are displayed in Fig.~\ref{fig:CIM}.